\documentclass{aastex7}
\usepackage{amsmath}
\usepackage{amssymb}
\usepackage{graphicx}
\newcommand{\middletablesize}{%
  \fontsize{6.8pt}{9pt}\selectfont
}

\begin{document}

\title{Gaia DR3 Reveals the Kinematics of Young Stellar Objects in the Carina Nebula: Peripheral Expansion and Central Rotation}

\author[orcid=0009-0007-2515-9199,gname=Xinyu, sname=Yang]{Xinyu Yang}
\affiliation{State Key Laboratory of Radio Astronomy and Technology, Xinjiang Astronomical Observatory, CAS, 150 Science 1-Street, Urumqi, Xinjiang, 830011, People's Republic of China}
\affiliation{University of Chinese Academy of Sciences, Beijing 100080, People's Republic of China}
\email{yangxinyu@xao.ac.cn}  

\author[orcid=0000-0002-8760-8988,gname=Yuxin, sname=He]{Yuxin He} 
\affiliation{State Key Laboratory of Radio Astronomy and Technology, Xinjiang Astronomical Observatory, CAS, 150 Science 1-Street, Urumqi, Xinjiang, 830011, People's Republic of China}
\affiliation{University of Chinese Academy of Sciences, Beijing 100080, People's Republic of China}
\affiliation{Xinjiang Key Laboratory of Radio Astrophysics, Urumqi 830011, People's Republic of China}
\email[show]{heyuxin@xao.ac.cn}
\correspondingauthor{Yuxin He}

\author[orcid=0000-0002-6820-198X,gname=Chaojie,sname=Hao]{Chaojie Hao}
\affiliation{Purple Mountain Observatory, Chinese Academy of Sciences, Nanjing 210023, People's Republic of China}
\email{cjhao@pmo.ac.cn}  

\author[orcid=0009-0000-5512-9109,gname=Yiwei,sname=Dong]{Yiwei Dong}
\affiliation{Purple Mountain Observatory, Chinese Academy of Sciences, Nanjing 210023, People's Republic of China}
\affiliation{School of Astronomy and Space Science, University of Science and Technology of China, Hefei 230026, People's Republic of China}
\email{dongyw@pmo.ac.cn}  

\author[orcid=0000-0001-5602-3306,gname=Ye,sname=Xu]{Ye Xu}
\affiliation{Purple Mountain Observatory, Chinese Academy of Sciences, Nanjing 210023, People's Republic of China}
\affiliation{School of Astronomy and Space Science, University of Science and Technology of China, Hefei 230026, People's Republic of China}
\affiliation{State Key Laboratory of Radio Astronomy and Technology, Xinjiang Astronomical Observatory, CAS, 150 Science 1-Street, Urumqi, Xinjiang, 830011, People's Republic of China}
\email{xuye@pmo.ac.cn}  

\author[orcid=0000-0001-7526-0120,gname=Yingjie,sname=Li]{Yingjie Li}
\affiliation{Purple Mountain Observatory, Chinese Academy of Sciences, Nanjing 210023, People's Republic of China}
\email{liyj@pmo.ac.cn}

\author[orcid=0009-0006-0392-6345,gname=Zehao,sname=Lin]{Zehao Lin}
\affiliation{Purple Mountain Observatory, Chinese Academy of Sciences, Nanjing 210023, People's Republic of China}
\email{linzh@pmo.ac.cn}  

\author[orcid=0009-0001-9837-9455,gname=Dejian,sname=Liu]{Dejian Liu}
\affiliation{College of Science, China Three Gorges University, Yichang 443000, People's Republic of China}
\email{liudejian@ctgu.edu.cn}

\author[orcid=0000-0003-4910-1390,gname=Jarken, sname=Esimbek]{Jarken Esimbek} 
\affiliation{State Key Laboratory of Radio Astronomy and Technology, Xinjiang Astronomical Observatory, CAS, 150 Science 1-Street, Urumqi, Xinjiang, 830011, People's Republic of China}
\affiliation{University of Chinese Academy of Sciences, Beijing 100080, People's Republic of China}
\affiliation{Xinjiang Key Laboratory of Radio Astrophysics, Urumqi 830011, People's Republic of China}
\email{jarken@xao.ac.cn}

\author[orcid=0000-0003-0356-818X,gname=Jianjun, sname=Zhou]{Jianjun Zhou} 
\affiliation{State Key Laboratory of Radio Astronomy and Technology, Xinjiang Astronomical Observatory, CAS, 150 Science 1-Street, Urumqi, Xinjiang, 830011, People's Republic of China}
\affiliation{University of Chinese Academy of Sciences, Beijing 100080, People's Republic of China}
\affiliation{Xinjiang Key Laboratory of Radio Astrophysics, Urumqi 830011, People's Republic of China}
\email{zhoujianjun@xao.ac.cn}

\author[orcid=0000-0003-0933-7112,gname=Gang, sname=Wu]{Gang Wu} 
\affiliation{State Key Laboratory of Radio Astronomy and Technology, Xinjiang Astronomical Observatory, CAS, 150 Science 1-Street, Urumqi, Xinjiang, 830011, People's Republic of China}
\affiliation{University of Chinese Academy of Sciences, Beijing 100080, People's Republic of China}
\affiliation{Xinjiang Key Laboratory of Radio Astrophysics, Urumqi 830011, People's Republic of China}
\email{wugang@xao.ac.cn}

\author[orcid=0000-0002-4154-4309,gname=Xindi, sname=Tang]{Xindi Tang} 
\affiliation{State Key Laboratory of Radio Astronomy and Technology, Xinjiang Astronomical Observatory, CAS, 150 Science 1-Street, Urumqi, Xinjiang, 830011, People's Republic of China}
\affiliation{University of Chinese Academy of Sciences, Beijing 100080, People's Republic of China}
\affiliation{Xinjiang Key Laboratory of Radio Astrophysics, Urumqi 830011, People's Republic of China}
\email{tangxindi@xao.ac.cn}

\author[orcid=0000-0001-5494-6238,gname=Dalei, sname=Li]{Dalei Li} 
\affiliation{State Key Laboratory of Radio Astronomy and Technology, Xinjiang Astronomical Observatory, CAS, 150 Science 1-Street, Urumqi, Xinjiang, 830011, People's Republic of China}
\affiliation{University of Chinese Academy of Sciences, Beijing 100080, People's Republic of China}
\affiliation{Xinjiang Key Laboratory of Radio Astrophysics, Urumqi 830011, People's Republic of China}
\email{lidalei@xao.ac.cn}

\begin{abstract}

We investigate the kinematics of the Carina Nebula using a sample of young stellar objects (YSOs) cross-matched with Gaia DR3, together with $^{12} \mathrm{CO}~(1\mbox{--}0)$ molecular gas observations and APOGEE radial-velocity data. 
The results show that the proper motions of the YSOs in the four peripheral subregions of Carina display a systematic outward expansion, and that the two-dimensional expansion velocities of the subregions are approximately \(0.57\mbox{--}4.56~\mathrm{km~s^{-1}}\). The YSOs in the Central Region exhibit ordered rotation around an axis at a position angle of \(86.9\pm1.5^\circ\), with a rotation velocity of \(1.75\pm0.03~\mathrm{km~s^{-1}}\).
Further momentum estimates suggest that the current expansion in the peripheral regions may result from the combined effects of feedback from massive OB stars and possible past supernova explosions.
These results indicate that feedback from massive stars plays a dominant role in the dynamical evolution of the giant molecular cloud in the Carina Nebula.

\end{abstract}

\keywords{\uat{Molecular clouds}{1072}---
\uat{Interstellar dynamics}{839}---
\uat{Interstellar molecules}{849}---
\uat{Star formation}{1569}}---


\section{Introduction}
\label{sec:intro}
Giant molecular clouds (MCs) and the stars formed within them are important constituents of galaxies. Giant MCs are the primary sites of star formation and feedback, and play a key role in the cycling of matter in the universe. Dense gas in molecular clouds (MCs) collapses to form massive stars, which continuously inject energy, momentum, and metals into the surrounding medium through feedback mechanisms such as photoionization, stellar winds, and subsequent supernova explosions (SNe). These feedback processes alter the kinematic properties and metal abundances of the surrounding gas, affect the formation of subsequent generations of stars, and drive the ongoing evolution of the interstellar medium (ISM) within galaxies \citep{Chevance2020}.

Characterizing the kinematics of MCs is essential for understanding their evolution and dissipation and also helps trace the evolutionary history of star formation within them. However, previous studies relying on molecular-line observations, such as CO emission, can obtain only the projected intensity distributions of MCs and gas radial velocities along the line of sight \citep{Nurnberger2002,Rebolledo2016}, but cannot obtain their transverse velocities perpendicular to the line of sight or their internal distance distribution. Young stellar objects (YSOs) inherit the kinematic properties of their parent MCs \citep{Lada1987,GuptaChen2022,Megeath2022}, making them ideal tracers of kinematics of MCs. Observationally, the velocity differences between YSOs and nearby molecular gas are typically within a few $\mathrm{km\,s^{-1}}$ \citep{Covey2006,Kounkel2017,Grossschedl2021,Yang2025}, indicating that the two are generally kinematically consistent.

In recent years, Gaia Data Release 3 (Gaia DR3;\citealt{GaiaCollaboration2023}) has provided high-precision astrometric parameters for more than 1.5 billion sources, including positions, parallaxes, and proper motions, thereby enabling the reconstruction of three-dimensional stellar kinematics. By combining molecular gas data with Gaia astrometric parameters and radial velocities (e.g., from the APOGEE survey; \citealt{Majewski2017}) of YSOs, we can reveal the morphology and kinematic properties of MCs and thus better understand their star formation histories \citep{Grossschedl2018,Grossschedl2021}. Kinematic studies based on Gaia proper motions have been applied to multiple star-forming regions and young stellar clusters, with recent work revealing systematic expansion or dispersal signatures in these environments \citep{CantatGaudin2019,Kuhn2019,Kuhn2020,MelnikDambis2020,Lim2023,Lim2025}. These kinematic signatures are commonly associated with gas dissipation, the internal dynamical evolution of stellar clusters, and feedback from massive stars. In addition, the kinematic properties of YSOs have been widely used to investigate the expansion, structural evolution, and turbulent processes of MCs \citep{Dzib2018,Kounkel2018,QuintanaWright2022,Ha2021,Zhou2022,Liu2024,Dong2024,Yang2025,Swiggum2025}.

The Carina Nebula (also known as NGC~3372) lies in the Carina--Sagittarius Arm of the Milky Way \citep{Vallee2014}, at a distance of approximately 2.3~kpc \citep{Smith2006MNRAS,GopplPreibisch2022}. It is one of the most prominent massive star-forming regions and \ion{H}{2} region complexes in the Milky Way \citep{Smith2000,Rebolledo2016}. The Carina Nebula hosts approximately 140 massive OB stars \citep{Alexander2016}, more than 1400 YSOs \citep{Povich2011}, three hydrogen-rich late-type Wolf--Rayet stars \citep{Smith2006MNRAS}.
One of them, WR~25 (HD~93162), is a recently confirmed hierarchical triple comprising WN6ha, O5(f$^{+}$), and O7 stars,
with an estimated combined mass of approximately $120\,M_\odot$
\citep{Gosset2026,Deshmukh2026}.
The region also hosts the luminous blue variable Eta Carinae (hereafter Eta Car; \citealt{KiminkiSmith2018}), whose initial mass is approximately $150\,M_\odot$ \citep{DavidsonHumphreys1997}. Intense feedback from massive stars in the Carina Nebula has shaped its central region into a bipolar bubble 
structure \citep{Smith2000}. The young stellar clusters Tr~14, Tr~15, and Tr~16 are located in this region and contain approximately $80\%$ of the massive stars in the entire Carina complex \citep{Roccatagliata2013}.
To date, numerous multiwavelength observations of the Carina Nebula have been conducted. Studies of CO and neutral carbon (\ion{C}{1}) line emission have revealed the complex structure, strong turbulent motions, and multiple velocity components of the molecular gas, suggesting that massive star formation may be 
associated with cloud--cloud collisions \citep{Zhang2001,Yonekura2005,Rebolledo2016,Fujita2021}. ATLASGAL and Herschel dust-continuum observations show that intense stellar radiation and stellar winds significantly alter the temperature and morphology of the clouds, dispersing the gas while potentially compressing dense clouds and triggering subsequent star formation \citep{Preibisch2011,Preibisch2012,Roccatagliata2013}. This region therefore serves as an ideal laboratory for investigating feedback effects during the formation of OB stellar populations.

In this work, we combine molecular gas data with Gaia astrometric parameters and APOGEE radial velocities for YSOs, and analyze the kinematic properties of YSOs in the Carina Nebula. This paper is organized as follows. Section~\ref{sec:data} describes the data and sample selection. Section~\ref{sec:results} presents the kinematic results for the Carina region, including the expansion of the peripheral regions and the rotational motion of the Central Region. Section~\ref{sec:discussion} discusses the possible feedback mechanisms driving the peripheral expansion. Finally, Section~\ref{sec:SUM} presents our conclusions.

\begin{figure*}[t!]
\begin{center}
\includegraphics[width=\textwidth]{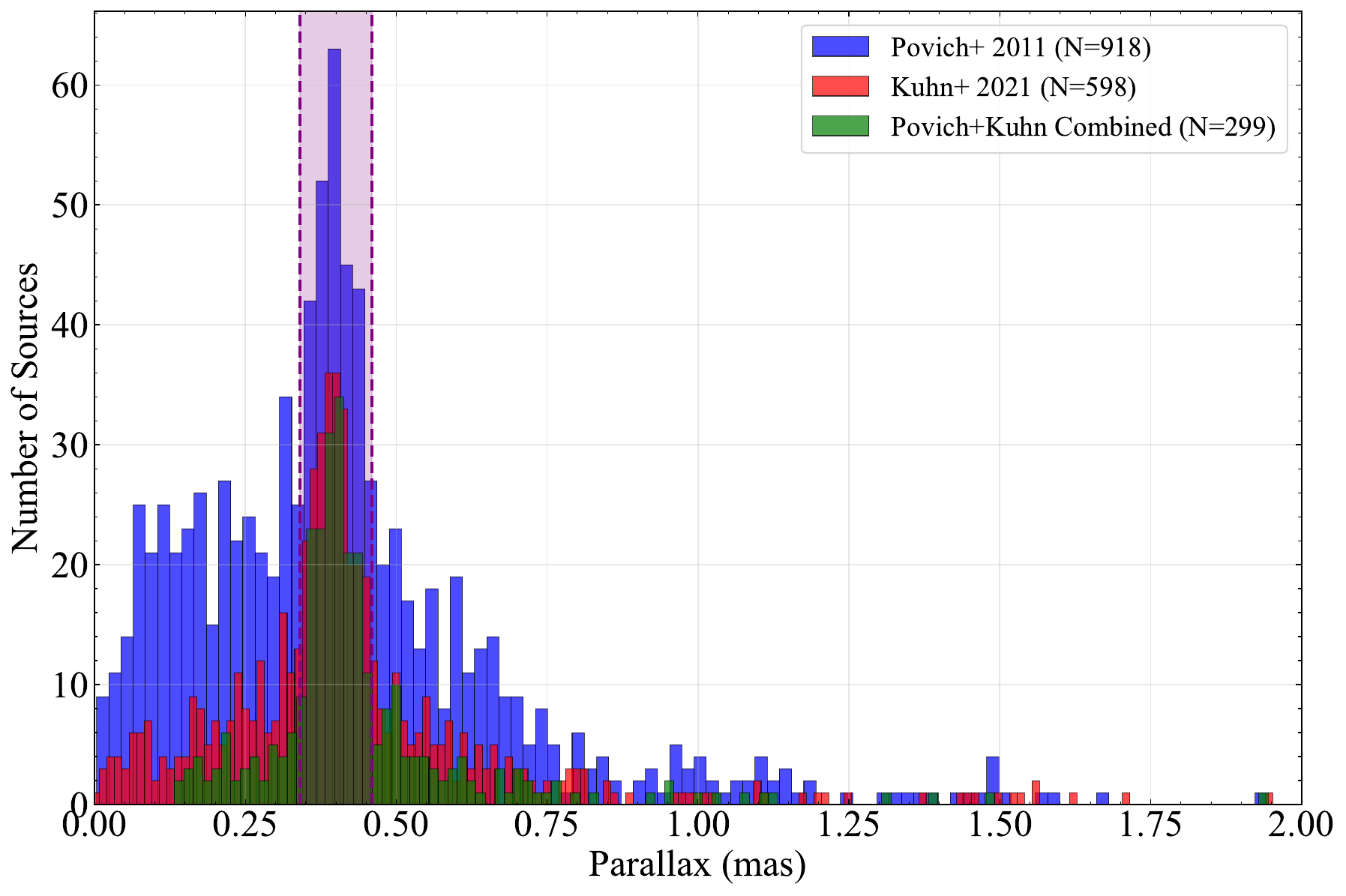}
\small
\caption{Parallax distribution of YSOs. The blue histogram (918 YSOs) and the red histogram (598 YSOs) show the parallax distributions of YSOs identified by \citet{Povich2011} and \citet{Kuhn2021}, respectively, after cross-matching with the Gaia DR3 catalog and retaining sources with $\mathrm{parallax}>0$. The green histogram represents the parallax distribution of the combined YSO sample after further selection (see Section~\ref{sec:yso_sample}). The purple dashed lines indicate the most densely populated and continuous parallax range ($0.34 < \mathrm{parallax} < 0.46$), which contains 170 YSOs.} 
\label{fig:parallax_distribution_overlay}
\end{center}
\end{figure*}

\section{Data} \label{sec:data}

\subsection{Molecular Line Data}
Carbon monoxide (CO) is the most widely used tracer of molecular gas. The $^{12}\mathrm{CO}~(1\mbox{--}0)$ molecular line data (115.3\,GHz) used in this study are taken from the Mopra Southern Galactic Plane CO Survey \citep{Burton2013,Braiding2015}. This survey is conducted using the Mopra 22\,m millimeter-wave telescope operated by the Australia Telescope National Facility of CSIRO. The telescope has an angular resolution of $33^{\prime\prime}$. At a velocity resolution of $0.088~\mathrm{km\,s^{-1}}$, the typical noise level is an antenna temperature of approximately $\sim1.6$\,K \citep{Rebolledo2016}, and the single-channel RMS noise of the data used in this work is $\sim1.5$\,K. The antenna brightness temperature is converted into the main-beam brightness temperature using the beam efficiency $\eta_{\mathrm{XB}}=0.55$ 
\citep{Ladd2005}.

\subsection{YSO Sample Data}
\label{sec:yso_sample}

The YSO sample in the Carina region used in this work is compiled from the catalogs published by \citet{Povich2011} and \citet{Kuhn2021}. The YSOs in both catalogs are identified and classified using infrared-excess techniques. Figure~\ref{fig:parallax_distribution_overlay} shows the parallax distribution of the YSOs. We first independently cross-match the YSOs identified by \citet{Povich2011} and \citet{Kuhn2021} with Gaia DR3 using a matching radius of \(1^{\prime\prime}\), retaining only sources with positive parallaxes (\(\mathrm{parallax}>0\)). This procedure yields 918 YSOs from \citet{Povich2011} (blue histogram) and 598 YSOs from \citet{Kuhn2021} (red histogram). We then merge the two Gaia-matched samples and remove duplicate sources, yielding a combined sample of 1043 unique YSOs.

To ensure the astrometric quality of the YSO sample and improve the reliability of the subsequent kinematic analysis, we apply the following selection criteria: (1) Renormalised Unit Weight Error ($\texttt{RUWE}$) $\leq1.4$; (2) $\texttt{parallax\_over\_error}>5$; and (3) proper-motion uncertainties that do not exceed the typical precision at \(G\simeq20\) mag, specifically $\sigma_{\mu_{\alpha *}}\leq0.487~\mathrm{mas~yr^{-1}}$ and $\sigma_{\mu_{\delta}}\leq0.428~\mathrm{mas~yr^{-1}}$ \citep{Lindegren2021}. This selection yields 299 YSOs that satisfy all the criteria(green histogram).
As shown in Figure~\ref{fig:parallax_distribution_overlay}, the parallax distribution of YSOs exhibits a prominent peak at approximately \(0.4~\mathrm{mas}\). We select the YSOs near this peak, with parallaxes between \(0.34\) and \(0.46~\mathrm{mas}\) (indicated by the purple dashed lines), as the initial sample for our analysis, yielding 170 YSOs. 

After correcting the parallaxes for the zero-point offset \citep{Lindegren2021}, we estimate the distance to the entire sample using a maximum-likelihood method \citep{CantatGaudin2018}, obtaining \(D=2310\pm10~\mathrm{pc}\). This estimate is consistent with the commonly adopted distance to the Carina Nebula of approximately \(2.3~\mathrm{kpc}\) \citep{GopplPreibisch2022}. The uncertainties in the above parameters are estimated using Monte Carlo simulations. Throughout the subsequent analysis, we adopt \(D=2.3~\mathrm{kpc}\).

\subsection{APOGEE Data}
\label{sec:apogee_data}

We adopted radial velocities from APOGEE Data Release 17 (DR17).
APOGEE provides high-resolution ($R\simeq22{,}500$) $H$-band spectra
covering $1.51$--$1.70~\mu{\rm m}$ \citep{Majewski2017}. The DR17
radial velocities were derived using the Doppler pipeline, which fits
all visit spectra simultaneously to determine the basic stellar
parameters and the radial velocity of each visit
\citep{Abdurrouf2022}. We used the cataloged mean radial velocity and
its uncertainty.
\par

\begin{figure*}[t!]
\begin{center}
\includegraphics[width=\textwidth]{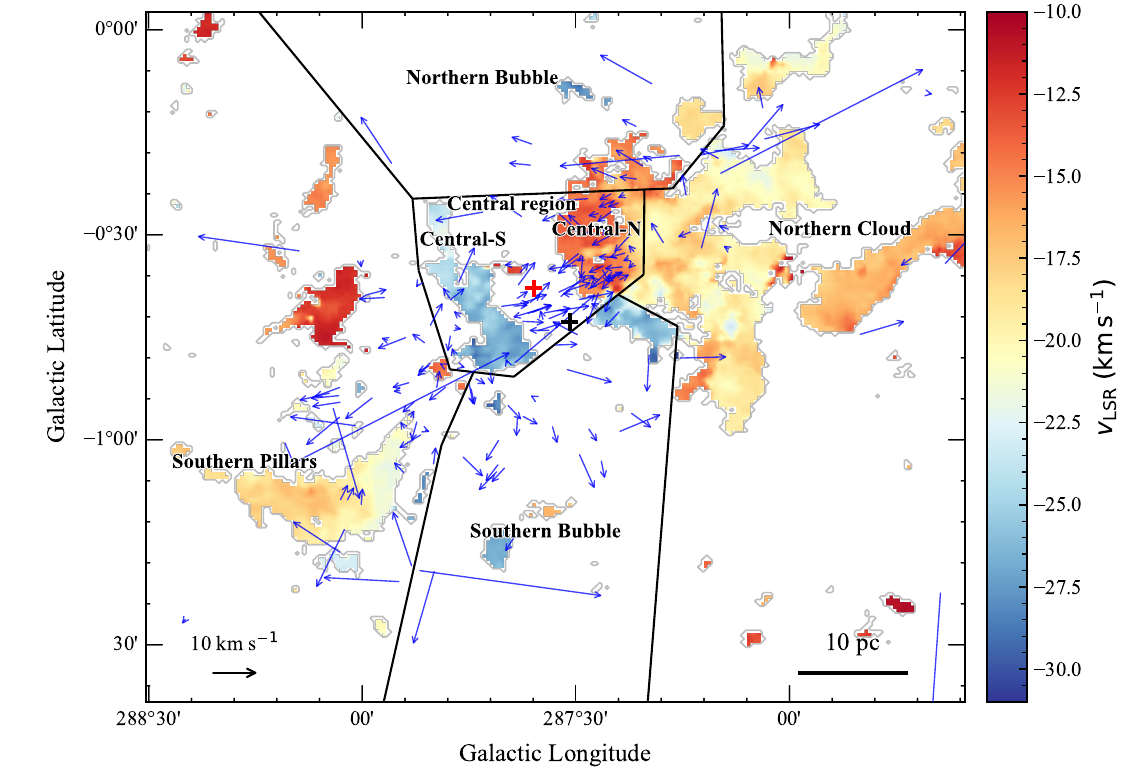}

\caption{Motions of YSOs in the Carina Nebula region. The residual tangential velocity of each YSO, calculated by subtracting the mean tangential velocity of the YSO sample from its individual tangential velocity, is represented by a blue arrow. The origin of each arrow marks the position of the corresponding YSO, while the direction and length of the arrow indicate the direction and magnitude of the residual tangential velocity, respectively. The arrows are overlaid on the $^{12}\mathrm{CO}$ moment-1 map. A scale of 10 pc is shown in the lower right corner of the image, and a scale of $5~\mathrm{km\,s^{-1}}$ is shown in the lower left corner. The red and black crosses mark the positions of Eta Car and WR~25, and the black lines delineate the boundaries of the five subregions.} 
\label{fig:figure2_right_regions}
\end{center}
\end{figure*}

\section{Methods and Results} \label{sec:results}
\subsection{Subregion Definition in the Carina Region}
\label{sec:overall_motion_of_the_Carina_Nebula_region_traced_by_YSOs}
 The molecular gas in the Carina Nebula has a highly nonuniform and complex distribution, with much of the high-column-density material located in filamentary, pillar-like, and arc-shaped structures around the periphery of the central ionized region \citep{Preibisch2012,Rebolledo2016}. In contrast, the molecular gas is relatively sparse in the Central Region around Eta Car, possibly because of the removal of the surrounding gas by feedback processes following massive star formation, including stellar radiation, stellar winds, and supernova explosions \citep{Smith2000,Rebolledo2016}. 

Figure~\ref{fig:figure2_right_regions} presents the line-of-sight velocity distribution of the molecular gas derived from the $^{12}\mathrm{CO}$ observations, together with the
spatial distribution and kinematic properties of the YSOs. The molecular gas in the Carina region spans a broad velocity range of approximately \(-30\) to \(-10~\mathrm{km~s^{-1}}\).
Previous CO observations reveal that the molecular gas in this region exhibits multiple velocity components and complex velocity gradients \citep{Rebolledo2016}. The YSOs are concentrated primarily near Eta Car in the Central Region of Carina and around the surrounding molecular gas structures. In peripheral regions where the molecular gas is relatively sparse, the number of YSOs decreases markedly.
Spitzer and Herschel observations have also shown that the YSO distribution in Carina is highly nonuniform, with many YSOs located near irradiated cloud edges and dust pillars rather than simply tracing the distribution of the total cloud mass \citep{Smith2010,Gaczkowski2013}.

Building on the regional division proposed by \citet{GopplPreibisch2022} based on dust, mid-infrared, and optical images, we divide the Carina region into five subregions by jointly considering the distribution of molecular gas in position--position--velocity (PPV) space and the relative proper motions of the YSOs. The five subregions are outlined by black lines in Figure~\ref{fig:figure2_right_regions}, where each of the four peripheral subregions connects to the Central Region. We further divide the Central Region into two parts based on the motions of the YSOs: the Central South Region and the Central North Region, hereafter referred to as Central-S and Central-N, respectively.

\begin{figure*}[t!]
\begin{center}
\includegraphics[width=\textwidth,angle=0]{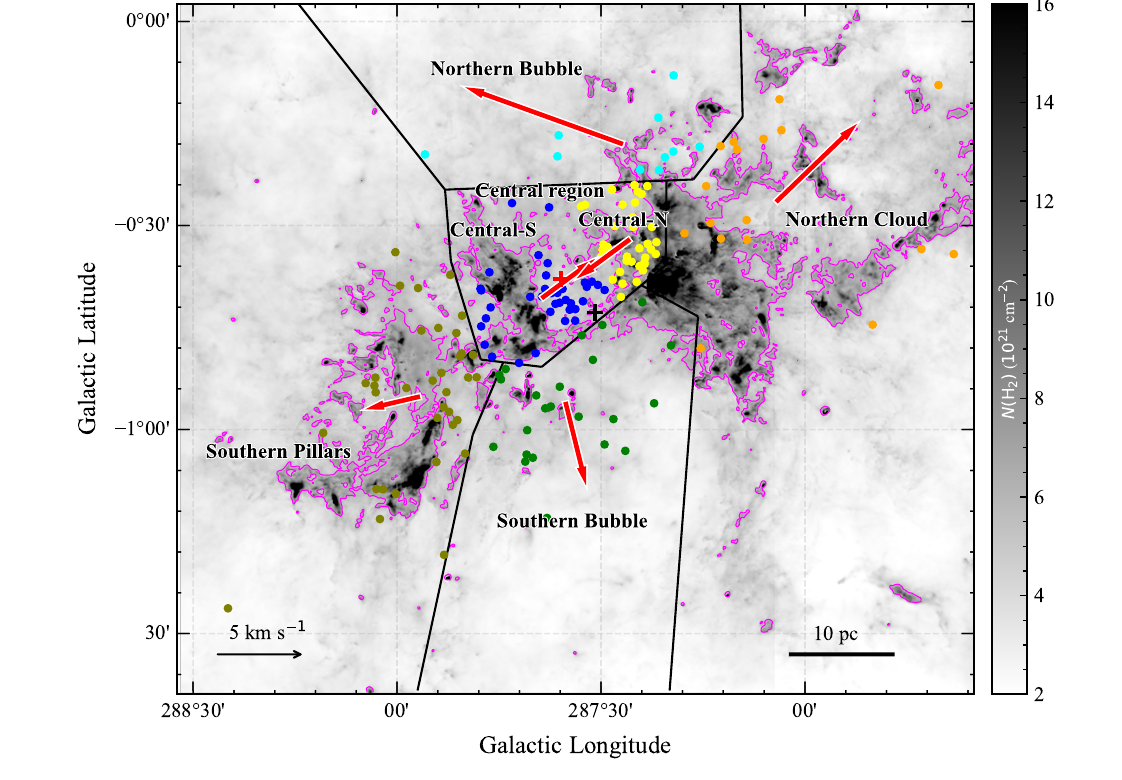}

\caption{The Carina Nebula is divided into the Central Region (Central-S and Central-N), Southern Pillars, Northern Cloud, Northern Bubble, and Southern Bubble subregions, which are delineated by black lines. The YSOs located in Central-S, Central-N, the Southern Pillars, the Northern Cloud, the Northern Bubble, and the Southern Bubble are marked by blue, yellow, olive, orange, cyan, and green circles, respectively.  The red and black crosses mark the positions of Eta Car and WR~25. The region boundaries and symbols are overlaid on the molecular hydrogen column-density map. The magenta contours mark a molecular hydrogen column density of $6\times10^{21}\,\mathrm{cm^{-2}}$. Red arrows indicate the relative proper motions of the six subregions, calculated by subtracting the mean proper motion of YSOs in the Central Region. The origin of each arrow marks the mean position of the YSOs in the corresponding subregion, while its direction and length represent the direction and magnitude of the relative proper motion, respectively. A scale of 10 pc is shown in the lower right corner of the image, and a scale of $5\,\mathrm{km\,s^{-1}}$ is shown in the lower left corner.} 
\label{fig:carina_six_regions_central_residual_vlb_3sigma_map}
\end{center}
\end{figure*}

\begin{figure}[t!]
\begin{center}
\includegraphics[height=8cm]{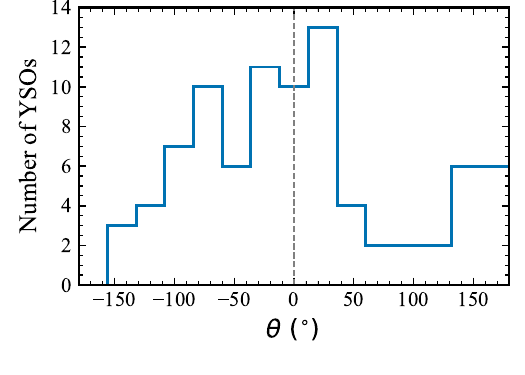}

\caption{Distribution of the angle $\theta$ between the relative proper-motion direction of each YSO in the four peripheral subregions and the corresponding radial direction relative to the reference center. The relative proper motions are obtained by subtracting the mean proper motion of the YSOs in the Central Region, and the geometric center of Central-S and Central-N is adopted as the reference center. Here, $\theta = 0^\circ$ indicates that the relative proper-motion direction of a YSO is exactly aligned with the radially outward direction from the reference center toward the YSO, while positive values indicate clockwise deviations from this radial direction.}
\label{fig:carina_six_regions_central_residual_vlb_3sigma_theta_histogram}
\end{center}
\end{figure}

\subsection{Systematic Expansion of the Peripheral Regions of Carina}
\label{sec:subregions_and_expansion}

Figure~\ref{fig:carina_six_regions_central_residual_vlb_3sigma_map} shows the relative motions of the five subregions. To investigate the motions of the surrounding subregions relative to the central feedback region, we subtract the mean proper motion of the YSOs in the Central Region, \((\mu_{l *},\mu_b)=(-6.99,-1.10)~\mathrm{mas~yr^{-1}}\). For each region, we exclude YSOs whose parallaxes or proper motions deviate from the corresponding mean values by more than \(3\sigma\). The origin of each red arrow marks the mean position of the YSOs in the corresponding subregion, while its direction and length represent the direction and magnitude of the relative motion, respectively.

Figure~\ref{fig:carina_six_regions_central_residual_vlb_3sigma_theta_histogram} shows the distribution of the angle $\theta$ between the proper-motion direction of each YSO in the four peripheral subregions and the radial direction from the reference center to the position of that YSO. To characterize the overall motion of the peripheral regions relative to the Central Region, we adopt the geometric center of Central-S and Central-N as the reference center. Figure~\ref{fig:carina_six_regions_central_residual_vlb_3sigma_theta_histogram} presents the distribution of \(\theta\) exhibits a pronounced excess near \(\theta=0^\circ\). Of the 86 peripheral YSOs included in the analysis, 39 satisfy \(\lvert\theta\rvert\leq45^\circ\), corresponding to \(45.35\%\) of the sample, while 44 satisfy \(\lvert\theta\rvert\leq60^\circ\), corresponding to \(51.16\%\). Both fractions exceed the respective values of \(25\%\) and \(33.33\%\) expected for an isotropic angular distribution. These results support a clear preference for radially outward motion among the peripheral YSOs, consistent with the kinematic picture of expansion in the peripheral regions.

For the four peripheral subregions, the magnitudes of their two-dimensional relative velocities are \(2.00\), \(5.01\), \(7.96\), and \(3.59~\mathrm{km~s^{-1}}\), respectively. By further projecting these tangential velocities onto the radial directions from the reference center toward the centers of the respective subregions, we obtain the corresponding expansion velocities of \(1.83\), \(4.56\), \(0.57\), and \(3.33~\mathrm{km~s^{-1}}\), respectively. The relative velocity components along Galactic longitude and latitude, observational parameters, and expansion velocities of the individual subregions are listed in Table~\ref{tab:proper_motion}. These results directly show that the molecular cloud in the Carina Nebula remains in an expanding state after several million years of stellar-wind feedback from the massive OB-star clusters in the Central Region. This finding supports the previous continuum-based inference that the Carina region exhibits expansion motions \citep{Preibisch2012} and is also consistent with the kinematic properties of Carina inferred by \citet{Kuhn2019} from YSOs in its stellar clusters.
\begingroup

\setlength{\tabcolsep}{1.5pt}
\renewcommand{\arraystretch}{1.1}

\begin{deluxetable*}{lccccccccccc}[t]

\tabletypesize{\middletablesize}
\tablewidth{\textwidth}
\tablecolumns{12}

\tablecaption{
Kinematic properties of the Carina subregions.
\label{tab:proper_motion}
}
\scriptsize
\tablehead{
\colhead{Subregion} &
\colhead{YSOs} &
\colhead{$l$} &
\colhead{$b$} &
\colhead{$\varpi$} &
\colhead{$\mu_{l}^{*}$} &
\colhead{$\mu_b$} &
\colhead{$v_{l,\mathrm{res}}$} &
\colhead{$v_{b,\mathrm{res}}$} &
\colhead{$v_{\rm exp,2D}$} &
\colhead{$M_{\rm tot,SFE}$} &
\colhead{$a_{\rm now}$}
\\[-1ex]
\colhead{} &
\colhead{} &
\colhead{(deg)} &
\colhead{(deg)} &
\colhead{(mas)} &
\colhead{(mas yr$^{-1}$)} &
\colhead{(mas yr$^{-1}$)} &
\colhead{(km s$^{-1}$)} &
\colhead{(km s$^{-1}$)} &
\colhead{(km s$^{-1}$)} &
\colhead{$(\times 10^{3}\,M_\odot)$} &
\colhead{(pc$^{2}$)}
\\[-1ex]
\colhead{(1)} &
\colhead{(2)} &
\colhead{(3)} &
\colhead{(4)} &
\colhead{(5)} &
\colhead{(6)} &
\colhead{(7)} &
\colhead{(8)} &
\colhead{(9)} &
\colhead{(10)} &
\colhead{(11)} &
\colhead{(12)}
}

\startdata
Southern Pillars& 36& 287.949& -0.920& 0.398($\pm$0.007)& -6.791($\pm$0.008)& -1.126($\pm$0.008)& 1.95($\pm$0.10)& -0.46($\pm$0.09)& 1.83($\pm$0.10)& 17.4& 207.25\\
Northern Cloud& 17& 287.067& -0.439& 0.395($\pm$0.008)& -7.300($\pm$0.009)& -0.765($\pm$0.009)& -3.60($\pm$0.14)& 3.47($\pm$0.14)& 4.56($\pm$0.15)& 48.1& 402.74\\
Northern Bubble& 10& 287.451& -0.299& 0.393($\pm$0.008)& -6.284($\pm$0.008)& -0.836($\pm$0.008)& 7.48($\pm$0.19)& 2.70($\pm$0.11)& 0.57($\pm$0.10)& 5.5& 55.62\\
Southern Bubble& 23& 287.586& -0.937& 0.398($\pm$0.008)& -7.049($\pm$0.010)& -1.404($\pm$0.009)& -0.86($\pm$0.12)& -3.49($\pm$0.13)& 3.33($\pm$0.13)& 4.5& 43.31\\
Central-S& 36& 287.641& -0.675& 0.400($\pm$0.005)& -7.157($\pm$0.006)& -0.946($\pm$0.006)& -2.04($\pm$0.06)& 1.50($\pm$0.06)& $\cdots$& 6.8& $\cdots$\\
Central-N& 36& 287.433& -0.537& 0.396($\pm$0.006)& -6.783($\pm$0.006)& -1.221($\pm$0.006)& 2.04($\pm$0.06)& -1.50($\pm$0.06)& $\cdots$& 9.1& $\cdots$\\
\enddata
\begin{flushleft}
\small
\textit{Note.}
Column (1) lists the subregion name, and Column (2) gives the number
of YSOs retained in each subregion. Columns (3)--(4) list the mean
Galactic coordinates $(l,b)$ of each subregion. Column (5) gives
the mean parallax $\varpi$. Columns (6) and (7) give the mean proper
motions $\mu_{l}^{*}$ and $\mu_b$, respectively, where
$\mu_{l}^{*}=\mu_{l}\cos b$. The uncertainties in the parallaxes
and proper motions in Columns (5)--(7) are propagated from Gaia observational uncertainties. Columns
(8) and (9) list the residual plane-of-sky velocity components along
Galactic longitude and latitude, respectively, obtained after
subtracting the mean proper motion of the combined Central-S and
Central-N reference sample. Column (10) gives the radial component of
the residual velocity projected along the direction from the adopted
expansion center to the mean position of each subregion; positive
values indicate outward motion. Column (11) gives the mass corrected
for the star formation efficiency, $M_{\rm tot,SFE}$. This mass is
estimated from the $^{12}\mathrm{CO}$ gas mass and the number of YSOs,
assuming a mean stellar mass of $0.5\,M_\odot$ and a star formation
efficiency of $\mathrm{SFE}=4.8\%$ \citep{Evans2009}. Column (12)
gives the projected area of the $^{12}\mathrm{CO}$  cloud, $a_{\rm now}$, assuming a
distance of $D=2.3~\mathrm{kpc}$.
\end{flushleft}
\end{deluxetable*}
\endgroup

\subsection{Rotation in the Central Region of Carina}
\label{sec:rotation_in_the_central_region}

Figure~\ref{fig:carina_six_regions_central_residual_vlb_3sigma_map} reveals that the mean proper motions of the YSOs in Central-S and
Central-N indicate a tendency for the two populations to move toward
each other, with each subregion having a projected two-dimensional velocity of \(5.06~\mathrm{km\,s^{-1}}\). The projected physical separation between the centers of Central-S and Central-N in the plane of the sky is approximately \(10~\mathrm{pc}\). Assuming that the two subregions approach each other at their velocity of \(5.06~\mathrm{km\,s^{-1}}\), the corresponding collision timescale is approximately \(1.94~\mathrm{Myr}\). The mean parallaxes of Central-S and Central-N are both approximately \(0.40~\mathrm{mas}\), and the two subregions can be regarded as being at the same distance within the uncertainties. Considering that the stellar clusters in the Central Region of Carina are already a few Myr old \citep{Preibisch2011,Wang2011}, this converging motion cannot be interpreted as a simple collision and may instead reflect more complex kinematic behavior in the Central Region.

To further investigate the kinematic properties of this region, we cross-match the YSOs with the APOGEE catalog, following the criteria described in Appendix~\ref{sec:kinematics_of_gas_and_YSOs}. Of these sources, 10 are located in Central-S and 18 in Central-N. The mean radial velocities of the two subregions are $-18.79~\mathrm{km\,s^{-1}}$ and $-15.03~\mathrm{km\,s^{-1}}$, respectively. Their mean tangential velocities are $(v_l,v_b)=(-2.04\pm0.06,\;1.50\pm0.06)~\mathrm{km\,s^{-1}}$ and $(v_l,v_b)=(2.04\pm0.06,\;-1.50\pm0.06)~\mathrm{km\,s^{-1}}$, respectively. The difference between the mean tangential velocities of the two subregions is approximately \(5.06~\mathrm{km\,s^{-1}}\). Meanwhile, the difference between their mean radial velocities is approximately \(3.76~\mathrm{km\,s^{-1}}\). The radial-velocity difference is comparable in magnitude to the tangential-velocity difference. Together with the locations of the two subregions on opposite sides of the Central Region and their oppositely directed tangential motions, this similarity suggests that the YSOs in the Central Region may exhibit coherent rotation.

\begin{figure*}[t!]
\begin{center}
\includegraphics[width=\textwidth]{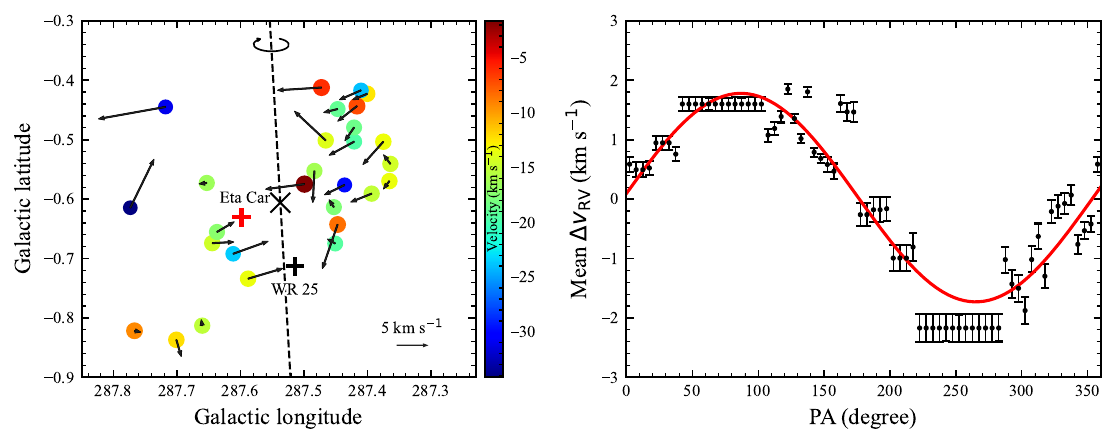}

\caption{Left: Spatial distribution of YSOs with radial-velocity measurements in the Central Region of the Carina Nebula. The circles are color-coded by radial velocity, while the black arrows indicate residual tangential velocities. The arrow in the lower right shows a scale of \(5\,\mathrm{km\,s^{-1}}\). The red and black crosses mark the positions of Eta Car and WR~25 respectively, while the black X marks the geometric center of Central-S and Central-N. The dashed line indicates the best-fitting rotation axis. The curved arrow at the top schematically indicates the
proposed direction of rotation. Right: The black circles show the mean residual radial velocity as a function of position angle (PA). The position angle of the rotation axis derived from the best-fitting sinusoidal curve is $86.9\pm1.5^\circ$, with a corresponding velocity amplitude of $1.75\pm0.03~\mathrm{km~s^{-1}}$. The black error bars show the uncertainties estimated using the Monte Carlo method, and the red curve represents the best-fitting sinusoidal function.}
\label{fig:final_combined}
\end{center}
\end{figure*}

To quantitatively test this interpretation, we follow the residual-velocity method adopted by \citet{LoktinPopov2020} and \citet{Hao2022}. Assuming that the rotation axis lies approximately in the plane of the sky, we search for a rotational signature by examining variations in the mean residual radial velocities of the YSOs on opposite sides of trial axes at different position angles. We adopt the geometric center of Central-S and Central-N, \((l_0,b_0)=(287.537^\circ,-0.606^\circ)\), as the kinematic center. We vary the position angle of a dividing line passing through the kinematic center from \(0^\circ\) to \(180^\circ\) in steps of \(5^\circ\). As the orientation of the dividing line changes, the YSOs assigned to the two sides are updated accordingly, and the mean radial velocities on the two sides are obtained as a function of position angle, as shown in the right panel of Figure~\ref{fig:final_combined}.
The fitting results show that the mean residual radial velocity varies approximately sinusoidally with position angle (PA). 
Here, PA is measured from the direction of increasing Galactic longitude (+l), taken as the reference direction, toward the direction of increasing Galactic latitude (+b), with positive angles defined in this direction. The best-fitting position angle of the rotation axis is \(86.9\pm1.5^\circ\), as indicated by the dashed line in the left panel of Figure~\ref{fig:final_combined}, with a corresponding projected rotational-velocity amplitude along the line of sight of \(1.75\pm0.03~\mathrm{km\,s^{-1}}\). Combined with the systematic differences between Central-S and Central-N in both radial- and tangential-velocity components, this result provides kinematic evidence for bulk rotation of the YSOs in the Central Region of Carina.

One possible interpretation of the rotational motion described above is that the YSOs in the Central Region inherited the angular momentum of their parent molecular cloud during their formation. Numerical simulations have shown that large-scale motions within MCs can generate coherent angular momentum on cloud scales and that this angular momentum can be partially preserved during star formation \citep{BurkertBodenheimer2000,Offner2008}. We therefore suggest that the rotational motion in the Central Region of Carina may be the kinematic outcome of the combined effects of the initial angular momentum of the parent molecular cloud, gas interactions, and feedback from massive stars.

\section{Possible Drivers of the Peripheral Expansion in Carina}
\label{sec:discussion}

The systematic outward motions observed for the YSOs in the peripheral regions of the Carina Nebula may be related to feedback from massive stars in the Central Region \citep{KiminkiSmith2018}. In the Carina Nebula, the main sources of feedback are massive stars concentrated in the Central Region, including Eta Car, WR~25, and Tr~16-244, along with other massive members of the young open clusters Tr~14, Tr~15, and Tr~16. The numerous OB stars in these clusters emit intense ionizing radiation (UV photons), which can produce large-scale \ion{H}{2} regions in the surrounding medium and drive the expansion of the gas \citep{Smith2006MNRAS}. 
Meanwhile, high-velocity stellar winds from massive stars can create bubble structures in the surrounding gas and further drive the outward dispersal of the molecular gas \citep{SmithBrooks2008}. The presence of the extremely massive star Eta Car and several massive stellar clusters of different ages in the Carina Nebula indicates that this region has undergone sustained massive-star formation and evolution. During its well-known Great Eruption in the nineteenth century, Eta Car injected substantial amounts of material and energy into its surroundings, significantly affecting the local ISM \citep{Davidson2001}.  WR~25 may also influence the surrounding ISM through its strong
stellar winds and ionizing radiation \citep{Smith2006MNRAS}. In addition, as some of the massive stars formed at earlier times evolve to late evolutionary stages, possible supernova explosions may inject additional energy and momentum into the surrounding medium and alter the structure and dynamical evolution of the molecular cloud \citep{WalchNaab2015}. The expansion of the molecular gas in the peripheral regions of Carina is therefore more likely to result from the long-term combined effects of multiple massive-star feedback processes, including ionizing radiation, stellar winds, and possible supernova explosions.

To further test whether the peripheral expansion identified in Section~\ref{sec:subregions_and_expansion} can be driven by feedback from massive stars in the Central Region of Carina, we estimate the expansion momentum required for the four peripheral subregions and compare it with the momentum that supernova explosions could transfer to these regions \citep{Grossschedl2021,Posch2023,Dong2024}. To derive the masses listed in Table~\ref{tab:proper_motion}, we adopt
$X_{\rm CO}=2.0\times10^{20}~
{\rm cm^{-2}\,(K\,km\,s^{-1})^{-1}}$
as the CO--${\rm H}_2$ conversion factor and convert the
$^{12}{\rm CO}$ integrated intensity into the molecular
hydrogen column density \citep{Bolatto2013}.
Following \citet{Konyves2020}, the molecular gas mass of the $i$th subregion
is then calculated as
\begin{equation}
M_i =
\delta A_{\rm pixel}\,
\mu_{{\rm H}_2}m_{\rm H}
\sum_{j\in i}N_j({\rm H}_2),
\label{eq:molecular_gas_mass}
\end{equation}where $\delta A_{\rm pixel}$ is the physical area corresponding to a single
pixel at an adopted distance of $D=2.3~{\rm kpc}$, $m_{\rm H}$ is the mass
of a hydrogen atom, and $\mu_{{\rm H}_2}=2.8$ is the mean molecular weight
per ${\rm H}_2$ molecule \citep{Kauffmann2008}. We assume a mean stellar mass of $0.5\,M_\odot$ and adopt a star formation 
efficiency of $\mathrm{SFE}=4.8\%$ \citep{Evans2009}. We obtain a total mass 
of approximately $7.55\times10^{4}\,M_\odot$ for the four peripheral regions 
of Carina, and their expansion momentum is
\begin{equation}
P_{\rm exp}
=
\sum_i M_i v_{\mathrm{exp,3D},i}
\simeq
3.30\times10^{5}\,
M_\odot\,\mathrm{km\,s^{-1}},
\label{eq:total_expansion_momentum}
\end{equation}which is assumed to be a conservative estimate of the expansion momentum. Here, \(M_i\) represents the mass of each subregion, and \(v_{{\rm exp},3D,i}\) represents the three-dimensional expansion velocity of each subregion. Since independent measurements of the expansion velocities along the line of sight are unavailable for individual subregions, we assume that the three-dimensional expansion velocity components are isotropically distributed and convert the two-dimensional expansion velocities (see Table~\ref{tab:proper_motion}) into three-dimensional velocities 
\citep{Angelo2020,Zhou2022}:
\begin{equation}
v_{\mathrm{exp,3D}}
\simeq
\sqrt{\frac{3}{2}}\,
v_{\mathrm{exp,2D}}.
\end{equation}

To estimate the feedback momentum that could be intercepted by the individual MCs, we approximate the four regions as spherical clouds. Their surface areas are estimated from their current projected sizes, while their initial surface areas remain uncertain. We therefore use the current projected $^{12}\mathrm{CO}$ areas of the four peripheral subregions, $a_{\rm now}$, listed in Table~\ref{tab:proper_motion}, and assume that their effective projected areas when affected by feedback are $(0.5\mbox{--}2)a_{\rm now}$ \citep{Grossschedl2021}. Numerical simulations show that, after entering the radiative phase, a typical core-collapse supernova can inject a momentum of approximately $P_{\rm SN}\simeq(2\mbox{--}4)\times10^{5}\,M_\odot\,\mathrm{km\,s^{-1}}$ into the surrounding ISM \citep{KimOstriker2015,IffrigHennebelle2015,WalchNaab2015}.
The effective momentum actually injected into the MCs depends on the explosion environment, cloud structure, and radiative cooling processes \citep{WalchNaab2015}. To estimate a lower limit, we assume that all of the supernova momentum within the solid angle $\Omega$ subtended by the four spherical regions is transferred to the MCs. Under the above assumptions, the momentum required to drive the observed molecular-cloud expansion is equivalent to that supplied by approximately three supernova explosions. This result is consistent with the inference by \citet{Hamaguchi2007}, based on the iron mass in the diffuse X-ray-emitting hot plasma of the Carina Nebula, that at least three and possibly as many as five supernova explosions had occurred. It is also consistent with the estimate of historical supernova activity in Tr~15 by \citet{Voss2012}. By combining the cluster age, the number of massive stars, and an $^{26}\mathrm{Al}$ production model, that study inferred that approximately three supernova explosions had previously occurred in Tr~15.

In addition to supernova feedback, we further examine whether continuous stellar winds from the massive stars in the Central Region can provide the expansion momentum required by the four peripheral MCs. The massive stars in Tr~14 and Tr~16 have ages of approximately \(1\)--\(3~\mathrm{Myr}\) \citep{Preibisch2011,Hur2012}, and we adopt \(\tau_{\rm w}=2~\mathrm{Myr}\) as the duration of continuous stellar feedback. We also adopt a representative stellar-wind velocity of \(v_{\rm w}=2000~\mathrm{km\,s^{-1}}\) for O- and B-type stars \citep{PrinjaMassa1998,Chen2013}. We again assume that the effective projected areas of the clouds when affected by feedback are \((0.5\mbox{--}2)a_{\rm now}\). Accounting for the fact that the four MCs intercept only the stellar winds within the solid angle \(\Omega\), we follow the methods of \citet{Arce2011} and \citet{DaleBonnell2008} and assume that the stellar-wind momentum intercepted by the MCs equals their expansion momentum. The stellar-wind mass-loss rate required to drive the expansion of the peripheral MCs is then
\begin{equation}
\dot M_{\rm w,req}
=
\frac{4\pi P_{\rm exp}}
{\Omega\,
v_{\rm w}\,\tau_{\rm w}}
=
(2.37\mbox{--}9.77)\times10^{-4}\,
M_\odot\,\mathrm{yr^{-1}}.
\end{equation}The typical stellar-wind mass-loss rate of an O- or B-type star is 
approximately \(\dot{M}_{\rm w,OB}\sim10^{-7}\mbox{--}10^{-6}\,M_\odot\,\mathrm{yr^{-1}}\) \citep{Chen2013}. The combined stellar-wind mass-loss rate of the approximately \(140\) massive OB stars currently confirmed in the Carina Nebula \citep{Alexander2016} is approximately \(10^{-5}\mbox{--}10^{-4}\,M_\odot\,\mathrm{yr^{-1}}\), which is comparable to the stellar-wind mass-loss rate calculated from our assumed model.

In summary, the present momentum estimates are subject to several sources of uncertainty and therefore cannot demonstrate that either supernova explosions or stellar winds alone can drive the expansion of the MCs in the peripheral regions of Carina. We therefore favor the following feedback scenario for the observed expansion of the peripheral regions of Carina: stellar winds, ionizing radiation, and radiation pressure from massive stars in the Carina Nebula accelerate the peripheral molecular gas over timescales of several Myr, while one or more past supernova explosions inject additional momentum into the MCs. Together, these processes produce the currently observed large-scale expansion. This combined-feedback scenario is consistent with previous studies of massive-star feedback in the Carina region \citep{Voss2012}.

\section{Summary}
\label{sec:SUM}

In this work, we combine a sample of YSOs with astrometric data from Gaia DR3, radial-velocity data from APOGEE, and the $^{12}\mathrm{CO}$ data from the Mopra telescope to investigate the kinematics of the MCs in the Carina Nebula. We systematically analyze the kinematic connection between the YSOs and their parent molecular gas. Based on the joint analysis of the spatial distributions and kinematic properties of YSOs and molecular gas, we further investigate the role of feedback from massive stars in regulating molecular cloud evolution and star formation processes. Our main results are summarized as follows:

\begin{enumerate}

\item After subtracting the mean proper motion of the Central Region, 
the YSOs in the Southern Pillars, Northern Cloud, Northern Bubble, 
and Southern Bubble show coherent radially outward motions. The corresponding projected expansion velocities of the four subregions are \(1.83\), \(4.56\), \(0.57\), and \(3.33~\mathrm{km\,s^{-1}}\), respectively.

\item The YSOs in the Central Region of Carina exhibit ordered rotation, with a rotation axis at a position angle of $86.9^\circ \pm 1.5^\circ$ and a rotational velocity of approximately $1.75 \pm 0.03\,\mathrm{km\,s^{-1}}$. This rotational motion may originate from angular momentum inherited from the parent molecular cloud during star formation. 

\item Our momentum analysis of the peripheral MCs in Carina indicates that continuous stellar winds from massive stars in the Central Region and past supernova explosions make important contributions to the momentum driving the molecular cloud expansion. We suggest that the observed expansion of the peripheral MCs is more likely to be driven by the combined effects of long-term stellar wind feedback and one or more past 
supernova explosions.
\end{enumerate}

In summary, using Gaia DR3 data, this study reveals the complex kinematic evolution of the Carina region: the peripheral YSOs exhibit systematic expansion, while the Central Region exhibits ordered rotation. These results support a scenario in which feedback from massive stars drives molecular cloud evolution and provide new observational constraints for understanding gas dispersal and kinematic evolution in massive star-forming regions. Future Gaia data releases will further improve the precision of three-dimensional kinematic measurements of YSOs and thereby provide deeper insight into how feedback shapes giant MCs.

\begin{acknowledgments}
This work was primarily funded by the National Key R\&D Program of China (Grant Nos. 2023YFA1608002 and 2022YFA1603103). It was also partially funded by the Xinjiang Talent Development Fund (grant No.XJRC-2025-KJ-YJ-CXPT-180), the Tianshan Talent Training Program (Grant No. 2024TSYCTD0013), the NSFC (Grant No. 12373029), the Chinese Academy of Sciences (CAS) “Light of West China” Program (Grant No. xbzg-zdsys-202212), the Natural Science Foundation of Xinjiang Uygur Autonomous Region (Grant Nos. 2026D01D58 and 2026D01E66), and the Central Guidance for Local Science and Technology Development Fund (Grant No. ZYYD2025ZY23). This work has made use of data from the European Space Agency (ESA) mission Gaia (https://www.cosmos.esa.int/gaia), processed by the Gaia Data Processing and Analysis Consortium (DPAC, https://www.cosmos.esa.int/web/gaia/dpac/consortium). Funding for the DPAC has been provided by national institutions, in particular, the institutions participating in the Gaia Multilateral Agreement.
\end{acknowledgments}

\software{Astropy \citep{Astropy2013,Astropy2018}, 
          Matplotlib \citep{Hunter2007}, 
          NumPy \citep{Harris2020}, 
          SAOImage DS9 \citep{JoyceMandel2003}, 
          TOPCAT \citep{Taylor2005},
          SciPy \citep{Virtanen2020}}


\appendix
\setcounter{figure}{0}
\renewcommand{\thefigure}{A\arabic{figure}}
\section{Kinematics of Gas and YSOs in the Central Region of the Carina Nebula}
\label{sec:kinematics_of_gas_and_YSOs}
\begin{figure}[h!]
\centering
\includegraphics[width=\textwidth]
{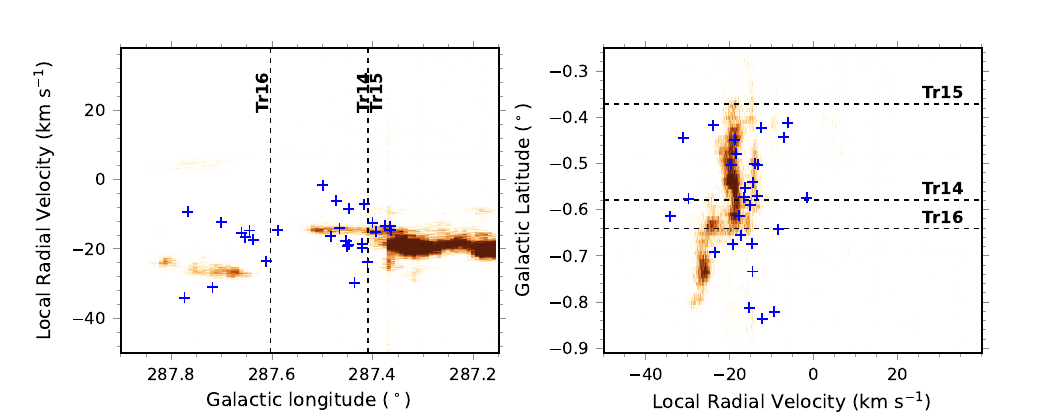}
\caption{Position--velocity diagrams of the
$^{12}\mathrm{CO}$ emission shown in pseudo-color, with
the selected YSOs overplotted as blue crosses. Left: $V_{\rm LSR}$ as
a function of Galactic longitude. Right: Galactic latitude as a
function of $V_{\rm LSR}$. The dashed lines mark the projected
positions of Tr~14, Tr~15, and Tr~16.}
\label{fig:Carina_12CO_PV_Glon_Glat_combined}
\end{figure}

Figure~\ref{fig:Carina_12CO_PV_Glon_Glat_combined} shows the Galactic
longitude--velocity and latitude--velocity diagrams of the
$^{12}\mathrm{CO}$ emission. We cross-match the YSOs
in this region with the APOGEE catalog, requiring an angular separation
of $<1^{\prime\prime}$, and obtain 55 matches. 
To facilitate comparison with the velocities of the CO gas, we convert the mean barycentric radial velocities provided by APOGEE for these YSOs to the local-standard-of-rest (LSR) frame and adopt the resulting velocities as $V_{\rm LSR}$. 
Among the cross-matched YSOs, we select 28 sources
satisfying $-35 < V_{\mathrm{LSR}} < 0\,\mathrm{km\,s^{-1}}$ for the subsequent kinematic analysis. This broad velocity interval encompasses the radial velocities associated with the molecular gas in Carina while excluding sources whose radial velocities are substantially offset from the CO emission. 
The 28 selected YSOs are overplotted on the $^{12}\mathrm{CO}$ position--velocity diagrams in
Figure~\ref{fig:Carina_12CO_PV_Glon_Glat_combined}.

For example, the YSOs located near Tr~14, the youngest stellar cluster in the Carina Nebula, are closely associated with the CO emission in both position and velocity. In contrast, the YSOs around the older stellar clusters Tr~15 and Tr~16 show larger offsets from the CO emission, particularly those surrounding Tr~16, which hosts the largest number of O-type stars.
Overall, the gas and YSOs in the Central Region of the Carina Nebula appear to be closely linked in both spatial distribution and kinematics. This result is consistent with the findings of \citet{KiminkiSmith2018} and suggests that the Carina Nebula formed from a single continuous molecular cloud that is now being accelerated and disrupted by the massive stars in the Central Region.


\bibliography{sample701}

@article{Alexander2016,
       author = {{Alexander}, Michael J. and {Hanes}, Richard J. and {Povich}, Matthew S. and {McSwain}, M. Virginia},
        title = "{A Catalog of New Spectroscopically Confirmed Massive OB Stars in Carina}",
      journal = {\aj},
         year = 2016,
        month = dec,
       volume = {152},
       number = {6},
          eid = {190},
        pages = {190},
          doi = {10.3847/0004-6256/152/6/190},
archivePrefix = {arXiv},
       eprint = {1611.01405},
 primaryClass = {astro-ph.SR},
       adsurl = {https://ui.adsabs.harvard.edu/abs/2016AJ....152..190A}
}

@article{Abdurrouf2022,
 author = {{Abdurro'uf} and {Accetta}, Katherine and {Aerts}, Conny and {Silva Aguirre}, V{\'\i}ctor and {Ahumada}, Romina and {Ajgaonkar}, Nikhil and {Filiz Ak}, N. and {Alam}, Shadab and {Allende Prieto}, Carlos and {Almeida}, Andr{\'e}s and {Anders}, Friedrich and {Anderson}, Scott F. and {Andrews}, Brett H. and {Anguiano}, Borja and {Aquino-Ort{\'\i}z}, Erik and {Arag{\'o}n-Salamanca}, Alfonso and {Argudo-Fern{\'a}ndez}, Maria and {Ata}, Metin and {Aubert}, Marie and {Avila-Reese}, Vladimir and {Badenes}, Carles and {Barb{\'a}}, Rodolfo H. and {Barger}, Kat and {Barrera-Ballesteros}, Jorge K. and {Beaton}, Rachael L. and {Beers}, Timothy C. and {Belfiore}, Francesco and {Bender}, Chad F. and {Bernardi}, Mariangela and {Bershady}, Matthew A. and {Beutler}, Florian and {Bidin}, Christian Moni and {Bird}, Jonathan C. and {Bizyaev}, Dmitry and {Blanc}, Guillermo A. and {Blanton}, Michael R. and {Boardman}, Nicholas Fraser and {Bolton}, Adam S. and {Boquien}, M{\'e}d{\'e}ric and {Borissova}, Jura and {Bovy}, Jo and {Brandt}, W.~N. and {Brown}, Jordan and {Brownstein}, Joel R. and {Brusa}, Marcella and {Buchner}, Johannes and {Bundy}, Kevin and {Burchett}, Joseph N. and {Bureau}, Martin and {Burgasser}, Adam and {Cabang}, Tuesday K. and {Campbell}, Stephanie and {Cappellari}, Michele and {Carlberg}, Joleen K. and {Wanderley}, F{\'a}bio Carneiro and {Carrera}, Ricardo and {Cash}, Jennifer and {Chen}, Yan-Ping and {Chen}, Wei-Huai and {Cherinka}, Brian and {Chiappini}, Cristina and {Choi}, Peter Doohyun and {Chojnowski}, S. Drew and {Chung}, Haeun and {Clerc}, Nicolas and {Cohen}, Roger E. and {Comerford}, Julia M. and {Comparat}, Johan and {da Costa}, Luiz and {Covey}, Kevin and {Crane}, Jeffrey D. and {Cruz-Gonzalez}, Irene and {Culhane}, Connor and {Cunha}, Katia and {Dai}, Y. Sophia and {Damke}, Guillermo and {Darling}, Jeremy and {Davidson}, Jr., James W. and {Davies}, Roger and {Dawson}, Kyle and {De Lee}, Nathan and {Diamond-Stanic}, Aleksandar M. and {Cano-D{\'\i}az}, Mariana and {S{\'a}nchez}, Helena Dom{\'\i}nguez and {Donor}, John and {Duckworth}, Chris and {Dwelly}, Tom and {Eisenstein}, Daniel J. and {Elsworth}, Yvonne P. and {Emsellem}, Eric and {Eracleous}, Mike and {Escoffier}, Stephanie and {Fan}, Xiaohui and {Farr}, Emily and {Feng}, Shuai and {Fern{\'a}ndez-Trincado}, Jos{\'e} G. and {Feuillet}, Diane and {Filipp}, Andreas and {Fillingham}, Sean P. and {Frinchaboy}, Peter M. and {Fromenteau}, Sebastien and {Galbany}, Llu{\'\i}s and {Garc{\'\i}a}, Rafael A. and {Garc{\'\i}a-Hern{\'a}ndez}, D.~A. and {Ge}, Junqiang and {Geisler}, Doug and {Gelfand}, Joseph and {G{\'e}ron}, Tobias and {Gibson}, Benjamin J. and {Goddy}, Julian and {Godoy-Rivera}, Diego and {Grabowski}, Kathleen and {Green}, Paul J. and {Greener}, Michael and {Grier}, Catherine J. and {Griffith}, Emily and {Guo}, Hong and {Guy}, Julien and {Hadjara}, Massinissa and {Harding}, Paul and {Hasselquist}, Sten and {Hayes}, Christian R. and {Hearty}, Fred and {Hern{\'a}ndez}, Jes{\'u}s and {Hill}, Lewis and {Hogg}, David W. and {Holtzman}, Jon A. and {Horta}, Danny and {Hsieh}, Bau-Ching and {Hsu}, Chin-Hao and {Hsu}, Yun-Hsin and {Huber}, Daniel and {Huertas-Company}, Marc and {Hutchinson}, Brian and {Hwang}, Ho Seong and {Ibarra-Medel}, H{\'e}ctor J. and {Chitham}, Jacob Ider and {Ilha}, Gabriele S. and {Imig}, Julie and {Jaekle}, Will and {Jayasinghe}, Tharindu and {Ji}, Xihan and {Johnson}, Jennifer A. and {Jones}, Amy and {J{\"o}nsson}, Henrik and {Katkov}, Ivan and {Khalatyan}, Dr., Arman and {Kinemuchi}, Karen and {Kisku}, Shobhit and {Knapen}, Johan H. and {Kneib}, Jean-Paul and {Kollmeier}, Juna A. and {Kong}, Miranda and {Kounkel}, Marina and {Kreckel}, Kathryn and {Krishnarao}, Dhanesh and {Lacerna}, Ivan and {Lane}, Richard R. and {Langgin}, Rachel and {Lavender}, Ramon and {Law}, David R. and {Lazarz}, Daniel and {Leung}, Henry W. and {Leung}, Ho-Hin and {Lewis}, Hannah M. and {Li}, Cheng and {Li}, Ran and {Lian}, Jianhui and {Liang}, Fu-Heng and {Lin}, Lihwai and {Lin}, Yen-Ting and {Lin}, Sicheng and {Lintott}, Chris and {Long}, Dan and {Longa-Pe{\~n}a}, Pen{\'e}lope and {L{\'o}pez-Cob{\'a}}, Carlos and {Lu}, Shengdong and {Lundgren}, Britt F. and {Luo}, Yuanze and {Mackereth}, J. Ted and {de la Macorra}, Axel and {Mahadevan}, Suvrath and {Majewski}, Steven R. and {Manchado}, Arturo and {Mandeville}, Travis and {Maraston}, Claudia and {Margalef-Bentabol}, Berta and {Masseron}, Thomas and {Masters}, Karen L. and {Mathur}, Savita and {McDermid}, Richard M. and {Mckay}, Myles and {Merloni}, Andrea and {Merrifield}, Michael and {Meszaros}, Szabolcs and {Miglio}, Andrea and {Di Mille}, Francesco and {Minniti}, Dante and {Minsley}, Rebecca and {Monachesi}, Antonela},
        title = "{The Seventeenth Data Release of the Sloan Digital Sky Surveys: Complete Release of MaNGA, MaStar, and APOGEE-2 Data}",
      journal = {\apjs},
         year = 2022,
        month = apr,
       volume = {259},
       number = {2},
          eid = {35},
        pages = {35},
          doi = {10.3847/1538-4365/ac4414},
archivePrefix = {arXiv},
       eprint = {2112.02026},
 primaryClass = {astro-ph.GA},
       adsurl = {https://ui.adsabs.harvard.edu/abs/2022ApJS..259...35A}
}

@article{Angelo2020,
       author = {{Angelo}, M.~S. and {Santos}, J.~F.~C. and {Corradi}, W.~J.~B.},
        title = "{Characterizing low-contrast Galactic open clusters with Gaia Data Release 2}",
      journal = {\mnras},
         year = 2020,
        month = apr,
       volume = {493},
       number = {3},
        pages = {3473-3489},
          doi = {10.1093/mnras/staa517},
archivePrefix = {arXiv},
       eprint = {2002.06158},
 primaryClass = {astro-ph.SR},
       adsurl = {https://ui.adsabs.harvard.edu/abs/2020MNRAS.493.3473A}
}

@article{Bolatto2013,
       author = {{Bolatto}, Alberto D. and {Wolfire}, Mark and {Leroy}, Adam K.},
        title = "{The CO-to-H$_{2}$ Conversion Factor}",
      journal = {\araa},
         year = 2013,
        month = aug,
       volume = {51},
       number = {1},
        pages = {207-268},
          doi = {10.1146/annurev-astro-082812-140944},
archivePrefix = {arXiv},
       eprint = {1301.3498},
 primaryClass = {astro-ph.GA},
       adsurl = {https://ui.adsabs.harvard.edu/abs/2013ARA&A..51..207B}
}

@article{Gosset2026,
       author = {{Gosset}, Eric and {Gamen}, Roberto and {Mahy}, Laurent and {Morrell}, Nidia and {Sana}, Hugues and {Barb{\'a}}, Rodolfo},
        title = "{A Detailed Spectroscopic Study of the WN6ha + O5 Colliding-wind Binary WR 25}",
      journal = {Bulletin de la Societe Royale des Sciences de Liege},
         year = 2026,
        month = jul,
       volume = {95},
        pages = {24-122},
          doi = {10.25518/0037-9565.12630},
archivePrefix = {arXiv},
       eprint = {2607.24390},
 primaryClass = {astro-ph.SR},
       adsurl = {https://ui.adsabs.harvard.edu/abs/2026BSRSL..95...24G}
}

@article{Deshmukh2026,
       author = {{Deshmukh}, K. and {Mahy}, L. and {Sana}, H. and {Frost}, A.~J. and {Gosset}, E. and {Lanthermann}, C. and {LeBouquin}, J.-B. and {Pauli}, D. and {Shenar}, T.},
        title = "{Southern massive stars at high angular resolution: WR 25 is a massive hierarchical triple system}",
      journal = {\aap},
         year = 2026,
        month = sep,
       volume = {713},
          eid = {A108},
        pages = {A108},
          doi = {10.1051/0004-6361/202661069},
archivePrefix = {arXiv},
       eprint = {2607.12836},
 primaryClass = {astro-ph.SR},
       adsurl = {https://ui.adsabs.harvard.edu/abs/2026A&A...713A.108D}
}

@article{Konyves2020,
       author = {{K{\"o}nyves}, V. and {Andr{\'e}}, Ph. and {Arzoumanian}, D. and {Schneider}, N. and {Men'shchikov}, A. and {Bontemps}, S. and {Ladjelate}, B. and {Didelon}, P. and {Pezzuto}, S. and {Benedettini}, M. and {Bracco}, A. and {Di Francesco}, J. and {Goodwin}, S. and {Rygl}, K.~L.~J. and {Shimajiri}, Y. and {Spinoglio}, L. and {Ward-Thompson}, D. and {White}, G.~J.},
        title = "{Properties of the dense core population in Orion B as seen by the Herschel Gould Belt survey}",
      journal = {\aap},
         year = 2020,
        month = mar,
       volume = {635},
          eid = {A34},
        pages = {A34},
          doi = {10.1051/0004-6361/201834753},
archivePrefix = {arXiv},
       eprint = {1910.04053},
 primaryClass = {astro-ph.SR},
       adsurl = {https://ui.adsabs.harvard.edu/abs/2020A&A...635A..34K}
}

@article{Kauffmann2008,
       author = {{Kauffmann}, J. and {Bertoldi}, F. and {Bourke}, T.~L. and {Evans}, II, N.~J. and {Lee}, C.~W.},
        title = "{MAMBO mapping of Spitzer c2d small clouds and cores}",
      journal = {\aap},
         year = 2008,
        month = sep,
       volume = {487},
       number = {3},
        pages = {993-1017},
          doi = {10.1051/0004-6361:200809481},
archivePrefix = {arXiv},
       eprint = {0805.4205},
 primaryClass = {astro-ph},
       adsurl = {https://ui.adsabs.harvard.edu/abs/2008A&A...487..993K}
}

@article{Braiding2015,
       author = {{Braiding}, Catherine and {Burton}, M.~G. and {Blackwell}, R. and {Gl{\"u}ck}, C. and {Hawkes}, J. and {Kulesa}, C. and {Maxted}, N. and {Rebolledo}, D. and {Rowell}, G. and {Stark}, A. and {Tothill}, N. and {Urquhart}, J.~S. and {Voisin}, F. and {Walsh}, A.~J. and {de Wilt}, P. and {Wong}, G.~F.},
        title = "{The Mopra Southern Galactic Plane CO Survey - Data Release 1}",
      journal = {\pasa},
         year = 2015,
        month = may,
       volume = {32},
          eid = {e020},
        pages = {e020},
          doi = {10.1017/pasa.2015.20},
archivePrefix = {arXiv},
       eprint = {1504.04068},
 primaryClass = {astro-ph.GA},
       adsurl = {https://ui.adsabs.harvard.edu/abs/2015PASA...32...20B}
}

@article{BurkertBodenheimer2000,
       author = {{Burkert}, Andreas and {Bodenheimer}, Peter},
        title = "{Turbulent Molecular Cloud Cores: Rotational Properties}",
      journal = {\apj},
         year = 2000,
        month = nov,
       volume = {543},
       number = {2},
        pages = {822-830},
          doi = {10.1086/317122},
archivePrefix = {arXiv},
       eprint = {astro-ph/0006010},
 primaryClass = {astro-ph},
       adsurl = {https://ui.adsabs.harvard.edu/abs/2000ApJ...543..822B}
}

@article{Burton2013,
       author = {{Burton}, Michael G. and {Braiding}, C. and {Glueck}, C. and {Goldsmith}, P. and {Hawkes}, J. and {Hollenbach}, D.~J. and {Kulesa}, C. and {Martin}, C.~L. and {Pineda}, J.~L. and {Rowell}, G. and {Simon}, R. and {Stark}, A.~A. and {Stutzki}, J. and {Tothill}, N.~J.~H. and {Urquhart}, J.~S. and {Walker}, C. and {Walsh}, A.~J. and {Wolfire}, M.},
        title = "{The Mopra Southern Galactic Plane CO Survey}",
      journal = {\pasa},
         year = 2013,
        month = aug,
       volume = {30},
          eid = {e044},
        pages = {e044},
          doi = {10.1017/pasa.2013.22},
archivePrefix = {arXiv},
       eprint = {1307.1027},
 primaryClass = {astro-ph.GA},
       adsurl = {https://ui.adsabs.harvard.edu/abs/2013PASA...30...44B}
}

@article{CantatGaudin2018,
       author = {{Cantat-Gaudin}, T. and {Jordi}, C. and {Vallenari}, A. and {Bragaglia}, A. and {Balaguer-N{\'u}{\~n}ez}, L. and {Soubiran}, C. and {Bossini}, D. and {Moitinho}, A. and {Castro-Ginard}, A. and {Krone-Martins}, A. and {Casamiquela}, L. and {Sordo}, R. and {Carrera}, R.},
        title = "{A Gaia DR2 view of the open cluster population in the Milky Way}",
      journal = {\aap},
         year = 2018,
        month = oct,
       volume = {618},
          eid = {A93},
        pages = {A93},
          doi = {10.1051/0004-6361/201833476},
archivePrefix = {arXiv},
       eprint = {1805.08726},
 primaryClass = {astro-ph.GA},
       adsurl = {https://ui.adsabs.harvard.edu/abs/2018A&A...618A..93C}
}

@article{CantatGaudin2019,
       author = {{Cantat-Gaudin}, T. and {Jordi}, C. and {Wright}, N.~J. and {Armstrong}, J.~J. and {Vallenari}, A. and {Balaguer-N{\'u}{\~n}ez}, L. and {Ramos}, P. and {Bossini}, D. and {Padoan}, P. and {Pelkonen}, V.~M. and {Mapelli}, M. and {Jeffries}, R.~D.},
        title = "{Expanding associations in the Vela-Puppis region. 3D structure and kinematics of the young population}",
      journal = {\aap},
         year = 2019,
        month = jun,
       volume = {626},
          eid = {A17},
        pages = {A17},
          doi = {10.1051/0004-6361/201834957},
archivePrefix = {arXiv},
       eprint = {1812.08114},
 primaryClass = {astro-ph.GA},
       adsurl = {https://ui.adsabs.harvard.edu/abs/2019A&A...626A..17C}
}

@article{Chevance2020,
       author = {{Chevance}, M{\'e}lanie and {Kruijssen}, J.~M. Diederik and {Vazquez-Semadeni}, Enrique and {Nakamura}, Fumitaka and {Klessen}, Ralf and {Ballesteros-Paredes}, Javier and {Inutsuka}, Shu-ichiro and {Adamo}, Angela and {Hennebelle}, Patrick},
        title = "{The Molecular Cloud Lifecycle}",
      journal = {\ssr},
         year = 2020,
        month = apr,
       volume = {216},
       number = {4},
          eid = {50},
        pages = {50},
          doi = {10.1007/s11214-020-00674-x},
archivePrefix = {arXiv},
       eprint = {2004.06113},
 primaryClass = {astro-ph.GA},
       adsurl = {https://ui.adsabs.harvard.edu/abs/2020SSRv..216...50C}
}

@article{Covey2006,
      author = {{Covey}, Kevin R. and {Greene}, Thomas P. and {Doppmann}, Greg W. and {Lada}, Charles J.},
        title = "{The Radial Velocity Distribution of Class I and Flat-Spectrum Protostars}",
      journal = {\aj},
         year = 2006,
        month = jan,
       volume = {131},
       number = {1},
        pages = {512-519},
          doi = {10.1086/498064},
archivePrefix = {arXiv},
       eprint = {astro-ph/0509264},
 primaryClass = {astro-ph},
       adsurl = {https://ui.adsabs.harvard.edu/abs/2006AJ....131..512C}
}

@article{DavidsonHumphreys1997,
       author = {{Davidson}, Kris and {Humphreys}, Roberta M.},
        title = "{Eta Carinae and Its Environment}",
      journal = {\araa},
         year = 1997,
        month = jan,
       volume = {35},
        pages = {1-32},
          doi = {10.1146/annurev.astro.35.1.1},
       adsurl = {https://ui.adsabs.harvard.edu/abs/1997ARA&A..35....1D}
}

@article{Davidson2001,
       author = {{Davidson}, Kris and {Smith}, Nathan and {Gull}, Theodore R. and {Ishibashi}, Kazunori and {Hillier}, D.~J.},
        title = "{The Shape and Orientation of the Homunculus Nebula Based on Spectroscopic Velocities}",
      journal = {\aj},
         year = 2001,
        month = mar,
       volume = {121},
       number = {3},
        pages = {1569-1577},
          doi = {10.1086/319419},
       adsurl = {https://ui.adsabs.harvard.edu/abs/2001AJ....121.1569D}
}

@article{Dong2024,
      author = {{Dong}, Yiwei and {Xu}, Ye and {Hao}, Chaojie and {Li}, Yingjie and {Liu}, Dejian and {Sun}, Yan and {Lin}, Zehao},
        title = "{3D Morphology and Motions of the Canis Major Region from Gaia DR3}",
      journal = {\aj},
         year = 2024,
        month = nov,
       volume = {168},
       number = {5},
          eid = {225},
        pages = {225},
          doi = {10.3847/1538-3881/ad77a8},
archivePrefix = {arXiv},
       eprint = {2409.01670},
 primaryClass = {astro-ph.GA},
       adsurl = {https://ui.adsabs.harvard.edu/abs/2024AJ....168..225D}
}

@article{Dzib2018,
       author = {{Dzib}, Sergio A. and {Loinard}, Laurent and {Ortiz-Le{\'o}n}, Gisela N. and {Rodr{\'\i}guez}, Luis F. and {Galli}, Phillip A.~B.},
        title = "{Distances and Kinematics of Gould Belt Star-forming Regions with Gaia DR2 Results}",
      journal = {\apj},
         year = 2018,
        month = nov,
       volume = {867},
       number = {2},
          eid = {151},
        pages = {151},
          doi = {10.3847/1538-4357/aae687},
archivePrefix = {arXiv},
       eprint = {1810.01917},
 primaryClass = {astro-ph.SR},
       adsurl = {https://ui.adsabs.harvard.edu/abs/2018ApJ...867..151D}
}

@article{Evans2009,
       author = {{Evans}, II, Neal J. and {Dunham}, Michael M. and {J{\o}rgensen}, Jes K. and {Enoch}, Melissa L. and {Mer{\'\i}n}, Bruno and {van Dishoeck}, Ewine F. and {Alcal{\'a}}, Juan M. and {Myers}, Philip C. and {Stapelfeldt}, Karl R. and {Huard}, Tracy L. and {Allen}, Lori E. and {Harvey}, Paul M. and {van Kempen}, Tim and {Blake}, Geoffrey A. and {Koerner}, David W. and {Mundy}, Lee G. and {Padgett}, Deborah L. and {Sargent}, Anneila I.},
        title = "{The Spitzer c2d Legacy Results: Star-Formation Rates and Efficiencies; Evolution and Lifetimes}",
      journal = {\apjs},
         year = 2009,
        month = apr,
       volume = {181},
       number = {2},
        pages = {321-350},
          doi = {10.1088/0067-0049/181/2/321},
archivePrefix = {arXiv},
       eprint = {0811.1059},
 primaryClass = {astro-ph},
       adsurl = {https://ui.adsabs.harvard.edu/abs/2009ApJS..181..321E}
}

@article{Fujita2021,
       author = {{Fujita}, Shinji and {Sano}, Hidetoshi and {Enokiya}, Rei and {Hayashi}, Katsuhiro and {Kohno}, Mikito and {Tsuge}, Kisetsu and {Tachihara}, Kengo and {Nishimura}, Atsushi and {Ohama}, Akio and {Yamane}, Yumiko and {Ohno}, Takahiro and {Yamada}, Rin and {Fukui}, Yasuo},
        title = "{Massive star formation in the Carina nebula complex and Gum 31. I. the Carina nebula complex}",
      journal = {\pasj},
         year = 2021,
        month = jan,
       volume = {73},
        pages = {S201-S219},
          doi = {10.1093/pasj/psaa078},
archivePrefix = {arXiv},
       eprint = {2003.13925},
 primaryClass = {astro-ph.GA},
       adsurl = {https://ui.adsabs.harvard.edu/abs/2021PASJ...73S.201F}
}

@article{GaiaCollaboration2023,
      author = {{Gaia Collaboration} and {Vallenari}, A. and {Brown}, A.~G.~A. and {Prusti}, T. and {de Bruijne}, J.~H.~J. and {Arenou}, F. and {Babusiaux}, C. and {Biermann}, M. and {Creevey}, O.~L. and {Ducourant}, C. and {Evans}, D.~W. and {Eyer}, L. and {Guerra}, R. and {Hutton}, A. and {Jordi}, C. and {Klioner}, S.~A. and {Lammers}, U.~L. and {Lindegren}, L. and {Luri}, X. and {Mignard}, F. and {Panem}, C. and {Pourbaix}, D. and {Randich}, S. and {Sartoretti}, P. and {Soubiran}, C. and {Tanga}, P. and {Walton}, N.~A. and {Bailer-Jones}, C.~A.~L. and {Bastian}, U. and {Drimmel}, R. and {Jansen}, F. and {Katz}, D. and {Lattanzi}, M.~G. and {van Leeuwen}, F. and {Bakker}, J. and {Cacciari}, C. and {Casta{\~n}eda}, J. and {De Angeli}, F. and {Fabricius}, C. and {Fouesneau}, M. and {Fr{\'e}mat}, Y. and {Galluccio}, L. and {Guerrier}, A. and {Heiter}, U. and {Masana}, E. and {Messineo}, R. and {Mowlavi}, N. and {Nicolas}, C. and {Nienartowicz}, K. and {Pailler}, F. and {Panuzzo}, P. and {Riclet}, F. and {Roux}, W. and {Seabroke}, G.~M. and {Sordo}, R. and {Th{\'e}venin}, F. and {Gracia-Abril}, G. and {Portell}, J. and {Teyssier}, D. and {Altmann}, M. and {Andrae}, R. and {Audard}, M. and {Bellas-Velidis}, I. and {Benson}, K. and {Berthier}, J. and {Blomme}, R. and {Burgess}, P.~W. and {Busonero}, D. and {Busso}, G. and {C{\'a}novas}, H. and {Carry}, B. and {Cellino}, A. and {Cheek}, N. and {Clementini}, G. and {Damerdji}, Y. and {Davidson}, M. and {de Teodoro}, P. and {Nu{\~n}ez Campos}, M. and {Delchambre}, L. and {Dell'Oro}, A. and {Esquej}, P. and {Fern{\'a}ndez-Hern{\'a}ndez}, J. and {Fraile}, E. and {Garabato}, D. and {Garc{\'\i}a-Lario}, P. and {Gosset}, E. and {Haigron}, R. and {Halbwachs}, J.-L. and {Hambly}, N.~C. and {Harrison}, D.~L. and {Hern{\'a}ndez}, J. and {Hestroffer}, D. and {Hodgkin}, S.~T. and {Holl}, B. and {Jan{\ss}en}, K. and {Jevardat de Fombelle}, G. and {Jordan}, S. and {Krone-Martins}, A. and {Lanzafame}, A.~C. and {L{\"o}ffler}, W. and {Marchal}, O. and {Marrese}, P.~M. and {Moitinho}, A. and {Muinonen}, K. and {Osborne}, P. and {Pancino}, E. and {Pauwels}, T. and {Recio-Blanco}, A. and {Reyl{\'e}}, C. and {Riello}, M. and {Rimoldini}, L. and {Roegiers}, T. and {Rybizki}, J. and {Sarro}, L.~M. and {Siopis}, C. and {Smith}, M. and {Sozzetti}, A. and {Utrilla}, E. and {van Leeuwen}, M. and {Abbas}, U. and {{\'A}brah{\'a}m}, P. and {Abreu Aramburu}, A. and {Aerts}, C. and {Aguado}, J.~J. and {Ajaj}, M. and {Aldea-Montero}, F. and {Altavilla}, G. and {{\'A}lvarez}, M.~A. and {Alves}, J. and {Anders}, F. and {Anderson}, R.~I. and {Anglada Varela}, E. and {Antoja}, T. and {Baines}, D. and {Baker}, S.~G. and {Balaguer-N{\'u}{\~n}ez}, L. and {Balbinot}, E. and {Balog}, Z. and {Barache}, C. and {Barbato}, D. and {Barros}, M. and {Barstow}, M.~A. and {Bartolom{\'e}}, S. and {Bassilana}, J.-L. and {Bauchet}, N. and {Becciani}, U. and {Bellazzini}, M. and {Berihuete}, A. and {Bernet}, M. and {Bertone}, S. and {Bianchi}, L. and {Binnenfeld}, A. and {Blanco-Cuaresma}, S. and {Blazere}, A. and {Boch}, T. and {Bombrun}, A. and {Bossini}, D. and {Bouquillon}, S. and {Bragaglia}, A. and {Bramante}, L. and {Breedt}, E. and {Bressan}, A. and {Brouillet}, N. and {Brugaletta}, E. and {Bucciarelli}, B. and {Burlacu}, A. and {Butkevich}, A.~G. and {Buzzi}, R. and {Caffau}, E. and {Cancelliere}, R. and {Cantat-Gaudin}, T. and {Carballo}, R. and {Carlucci}, T. and {Carnerero}, M.~I. and {Carrasco}, J.~M. and {Casamiquela}, L. and {Castellani}, M. and {Castro-Ginard}, A. and {Chaoul}, L. and {Charlot}, P. and {Chemin}, L. and {Chiaramida}, V. and {Chiavassa}, A. and {Chornay}, N. and {Comoretto}, G. and {Contursi}, G. and {Cooper}, W.~J. and {Cornez}, T. and {Cowell}, S. and {Crifo}, F. and {Cropper}, M. and {Crosta}, M. and {Crowley}, C. and {Dafonte}, C. and {Dapergolas}, A. and {David}, M. and {David}, P. and {de Laverny}, P. and {De Luise}, F. and {De March}, R.},
        title = "{Gaia Data Release 3. Summary of the content and survey properties}",
      journal = {\aap},
         year = 2023,
        month = jun,
       volume = {674},
          eid = {A1},
        pages = {A1},
          doi = {10.1051/0004-6361/202243940},
archivePrefix = {arXiv},
       eprint = {2208.00211},
 primaryClass = {astro-ph.GA},
       adsurl = {https://ui.adsabs.harvard.edu/abs/2023A&A...674A...1G}
}

@ARTICLE{Gaczkowski2013,
       author = {{Gaczkowski}, B. and {Preibisch}, T. and {Ratzka}, T. and {Roccatagliata}, V. and {Ohlendorf}, H. and {Zinnecker}, H.},
        title = "{Herschel far-infrared observations of the Carina Nebula complex. II. The embedded young stellar and protostellar population}",
      journal = {\aap},
         year = 2013,
        month = jan,
       volume = {549},
          eid = {A67},
        pages = {A67},
          doi = {10.1051/0004-6361/201219836},
archivePrefix = {arXiv},
       eprint = {1211.2995},
 primaryClass = {astro-ph.GA},
       adsurl = {https://ui.adsabs.harvard.edu/abs/2013A&A...549A..67G}
}

@article{GopplPreibisch2022,
       author = {{G{\"o}ppl}, C. and {Preibisch}, T.},
        title = "{Gaia EDR3 distances of the young stellar clusters in the extended Carina Nebula complex}",
      journal = {\aap},
         year = 2022,
        month = apr,
       volume = {660},
          eid = {A11},
        pages = {A11},
          doi = {10.1051/0004-6361/202142576},
archivePrefix = {arXiv},
       eprint = {2201.09097},
 primaryClass = {astro-ph.SR},
       adsurl = {https://ui.adsabs.harvard.edu/abs/2022A&A...660A..11G}
}

@article{Grossschedl2018,
       author = {{Gro{\ss}schedl}, Josefa E. and {Alves}, Jo{\~a}o and {Meingast}, Stefan and {Ackerl}, Christine and {Ascenso}, Joana and {Bouy}, Herv{\'e} and {Burkert}, Andreas and {Forbrich}, Jan and {F{\"u}rnkranz}, Verena and {Goodman}, Alyssa and {Hacar}, {\'A}lvaro and {Herbst-Kiss}, Gabor and {Lada}, Charles J. and {Larreina}, Irati and {Leschinski}, Kieran and {Lombardi}, Marco and {Moitinho}, Andr{\'e} and {Mortimer}, Daniel and {Zari}, Eleonora},
        title = "{3D shape of Orion A from Gaia DR2}",
      journal = {\aap},
         year = 2018,
        month = nov,
       volume = {619},
          eid = {A106},
        pages = {A106},
          doi = {10.1051/0004-6361/201833901},
archivePrefix = {arXiv},
       eprint = {1808.05952},
 primaryClass = {astro-ph.GA},
       adsurl = {https://ui.adsabs.harvard.edu/abs/2018A&A...619A.106G}
}

@article{Grossschedl2021,
       author = {{Gro{\ss}schedl}, Josefa E. and {Alves}, Jo{\~a}o and {Meingast}, Stefan and {Herbst-Kiss}, Gabor},
        title = "{3D dynamics of the Orion cloud complex. Discovery of coherent radial gas motions at the 100-pc scale}",
      journal = {\aap},
         year = 2021,
        month = mar,
       volume = {647},
          eid = {A91},
        pages = {A91},
          doi = {10.1051/0004-6361/202038913},
archivePrefix = {arXiv},
       eprint = {2007.07254},
 primaryClass = {astro-ph.SR},
       adsurl = {https://ui.adsabs.harvard.edu/abs/2021A&A...647A..91G}
}

@article{GuptaChen2022,
       author = {{Gupta}, Aashish and {Chen}, Wen-Ping},
        title = "{Interplay between Young Stars and Molecular Clouds in the Ophiuchus Star-forming Complex}",
      journal = {\aj},
         year = 2022,
        month = may,
       volume = {163},
       number = {5},
          eid = {233},
        pages = {233},
          doi = {10.3847/1538-3881/ac5cc8},
archivePrefix = {arXiv},
       eprint = {2204.13797},
 primaryClass = {astro-ph.SR},
       adsurl = {https://ui.adsabs.harvard.edu/abs/2022AJ....163..233G}
}

@article{Posch2023,
       author = {{Posch}, Laura and {Miret-Roig}, N{\'u}ria and {Alves}, Jo{\~a}o and {Ratzenb{\"o}ck}, Sebastian and {Gro{\ss}schedl}, Josefa and {Meingast}, Stefan and {Zucker}, Catherine and {Burkert}, Andreas},
        title = "{The Corona Australis star formation complex is accelerating away from the Galactic plane}",
      journal = {\aap},
         year = 2023,
        month = nov,
       volume = {679},
          eid = {L10},
        pages = {L10},
          doi = {10.1051/0004-6361/202347186},
archivePrefix = {arXiv},
       eprint = {2310.14373},
 primaryClass = {astro-ph.GA},
       adsurl = {https://ui.adsabs.harvard.edu/abs/2023A&A...679L..10P}
}

@article{IffrigHennebelle2015,
       author = {{Iffrig}, Olivier and {Hennebelle}, Patrick},
        title = "{Mutual influence of supernovae and molecular clouds}",
      journal = {\aap},
         year = 2015,
        month = apr,
       volume = {576},
          eid = {A95},
        pages = {A95},
          doi = {10.1051/0004-6361/201424556},
archivePrefix = {arXiv},
       eprint = {1410.7972},
 primaryClass = {astro-ph.GA},
       adsurl = {https://ui.adsabs.harvard.edu/abs/2015A&A...576A..95I}
}

@article{WalchNaab2015,
       author = {{Walch}, Stefanie and {Naab}, Thorsten},
        title = "{The energy and momentum input of supernova explosions in structured and ionized molecular clouds}",
      journal = {\mnras},
         year = 2015,
        month = aug,
       volume = {451},
       number = {3},
        pages = {2757-2771},
          doi = {10.1093/mnras/stv1155},
archivePrefix = {arXiv},
       eprint = {1410.0011},
 primaryClass = {astro-ph.GA},
       adsurl = {https://ui.adsabs.harvard.edu/abs/2015MNRAS.451.2757W}
}

@article{Hamaguchi2007,
       author = {{Hamaguchi}, Kenji and {Petre}, Robert and {Matsumoto}, Hironori and {Tsujimoto}, Masahiro and {Holt}, Stephen S. and {Ezoe}, Yuichiro and {Ozawa}, Hideki and {Tsuboi}, Yohko and {Soong}, Yang and {Kitamoto}, Shunji and {Sekiguchi}, Akiko and {Kokubun}, Motohide},
        title = "{Suzaku Observation of Diffuse X-Ray Emission from the Carina Nebula}",
      journal = {\pasj},
         year = 2007,
        month = jan,
       volume = {59},
        pages = {151-161},
          doi = {10.1093/pasj/59.sp1.S151},
archivePrefix = {arXiv},
       eprint = {astro-ph/0610024},
 primaryClass = {astro-ph},
       adsurl = {https://ui.adsabs.harvard.edu/abs/2007PASJ...59S.151H}
}

@article{Voss2012,
       author = {{Voss}, R. and {Martin}, P. and {Diehl}, R. and {Vink}, J.~S. and {Hartmann}, D.~H. and {Preibisch}, T.},
        title = "{Energetic feedback and $^{26}$Al from massive stars and their supernovae in the Carina region}",
      journal = {\aap},
         year = 2012,
        month = mar,
       volume = {539},
          eid = {A66},
        pages = {A66},
          doi = {10.1051/0004-6361/201118209},
archivePrefix = {arXiv},
       eprint = {1202.0282},
 primaryClass = {astro-ph.GA},
       adsurl = {https://ui.adsabs.harvard.edu/abs/2012A&A...539A..66V}
}

@article{Preibisch2011,
       author = {{Preibisch}, Thomas and {Hodgkin}, Simon and {Irwin}, Mike and {Lewis}, James R. and {King}, Robert R. and {McCaughrean}, Mark J. and {Zinnecker}, Hans and {Townsley}, Leisa and {Broos}, Patrick},
        title = "{Near-infrared Properties of the X-ray-emitting Young Stellar Objects in the Carina Nebula}",
      journal = {\apjs},
         year = 2011,
        month = may,
       volume = {194},
       number = {1},
          eid = {10},
        pages = {10},
          doi = {10.1088/0067-0049/194/1/10},
archivePrefix = {arXiv},
       eprint = {1103.2052},
 primaryClass = {astro-ph.SR},
       adsurl = {https://ui.adsabs.harvard.edu/abs/2011ApJS..194...10P}
}

@article{Hur2012,
       author = {{Hur}, Hyeonoh and {Sung}, Hwankyung and {Bessell}, Michael S.},
        title = "{Distance and the Initial Mass Function of Young Open Clusters in the {\ensuremath{\eta}} Carina Nebula: Tr 14 and Tr 16}",
      journal = {\aj},
         year = 2012,
        month = feb,
       volume = {143},
       number = {2},
          eid = {41},
        pages = {41},
          doi = {10.1088/0004-6256/143/2/41},
archivePrefix = {arXiv},
       eprint = {1201.0623},
 primaryClass = {astro-ph.SR},
       adsurl = {https://ui.adsabs.harvard.edu/abs/2012AJ....143...41H}
}

@article{PrinjaMassa1998,
       author = {{Prinja}, Raman K. and {Massa}, Derck and {Howarth}, Ian D. and {Fullerton}, Alex W.},
        title = "{Repetitive structure in the stellar wind of HD 93843: a normal O-type star}",
      journal = {\mnras},
         year = 1998,
        month = dec,
       volume = {301},
       number = {4},
        pages = {926-934},
          doi = {10.1046/j.1365-8711.1998.01984.x},
       adsurl = {https://ui.adsabs.harvard.edu/abs/1998MNRAS.301..926P}
}

@article{Chen2013,
       author = {{Chen}, Yang and {Zhou}, Ping and {Chu}, You-Hua},
        title = "{Linear Relation for Wind-blown Bubble Sizes of Main-sequence OB Stars in a Molecular Environment and Implication for Supernova Progenitors}",
      journal = {\apjl},
         year = 2013,
        month = may,
       volume = {769},
       number = {1},
          eid = {L16},
        pages = {L16},
          doi = {10.1088/2041-8205/769/1/L16},
archivePrefix = {arXiv},
       eprint = {1304.5126},
 primaryClass = {astro-ph.GA},
       adsurl = {https://ui.adsabs.harvard.edu/abs/2013ApJ...769L..16C}
}

@article{Arce2011,
       author = {{Arce}, H{\'e}ctor G. and {Borkin}, Michelle A. and {Goodman}, Alyssa A. and {Pineda}, Jaime E. and {Beaumont}, Christopher N.},
        title = "{A Bubbling Nearby Molecular Cloud: COMPLETE Shells in Perseus}",
      journal = {\apj},
         year = 2011,
        month = dec,
       volume = {742},
       number = {2},
          eid = {105},
        pages = {105},
          doi = {10.1088/0004-637X/742/2/105},
archivePrefix = {arXiv},
       eprint = {1109.3368},
 primaryClass = {astro-ph.SR},
       adsurl = {https://ui.adsabs.harvard.edu/abs/2011ApJ...742..105A}
}

@article{DaleBonnell2008,
       author = {{Dale}, J.~E. and {Bonnell}, I.~A.},
        title = "{The effect of stellar winds on the formation of a protocluster}",
      journal = {\mnras},
         year = 2008,
        month = nov,
       volume = {391},
       number = {1},
        pages = {2-13},
          doi = {10.1111/j.1365-2966.2008.13802.x},
archivePrefix = {arXiv},
       eprint = {0808.1510},
 primaryClass = {astro-ph},
       adsurl = {https://ui.adsabs.harvard.edu/abs/2008MNRAS.391....2D}
}

@article{Ha2021,
       author = {{Ha}, Trung and {Li}, Yuan and {Xu}, Siyao and {Kounkel}, Marina and {Li}, Hui},
        title = "{Measuring Turbulence with Young Stars in the Orion Complex}",
      journal = {\apjl},
         year = 2021,
        month = feb,
       volume = {907},
       number = {2},
          eid = {L40},
        pages = {L40},
          doi = {10.3847/2041-8213/abd8c9},
archivePrefix = {arXiv},
       eprint = {2101.03176},
 primaryClass = {astro-ph.GA},
       adsurl = {https://ui.adsabs.harvard.edu/abs/2021ApJ...907L..40H}
}

@article{Hao2022,
       author = {{Hao}, C.~J. and {Xu}, Y. and {Bian}, S.~B. and {Hou}, L.~G. and {Lin}, Z.~H. and {Li}, Y.~J. and {Liu}, D.~J.},
        title = "{On the Nature of Rotation in the Praesepe Cluster}",
      journal = {\apj},
         year = 2022,
        month = oct,
       volume = {938},
       number = {2},
          eid = {100},
        pages = {100},
          doi = {10.3847/1538-4357/ac92fc},
archivePrefix = {arXiv},
       eprint = {2210.01420},
 primaryClass = {astro-ph.GA},
       adsurl = {https://ui.adsabs.harvard.edu/abs/2022ApJ...938..100H}
}

@ARTICLE{Hunter2007,
       author = {{Hunter}, John D.},
        title = "{Matplotlib: A 2D Graphics Environment}",
      journal = {Computing in Science and Engineering},
         year = 2007,
        month = jan,
       volume = {9},
       number = {3},
        pages = {90-95},
          doi = {10.1109/MCSE.2007.55},
       adsurl = {https://ui.adsabs.harvard.edu/abs/2007CSE.....9...90H}
}

@article{KimOstriker2015,
       author = {{Kim}, Chang-Goo and {Ostriker}, Eve C.},
        title = "{Momentum Injection by Supernovae in the Interstellar Medium}",
      journal = {\apj},
         year = 2015,
        month = apr,
       volume = {802},
       number = {2},
          eid = {99},
        pages = {99},
          doi = {10.1088/0004-637X/802/2/99},
archivePrefix = {arXiv},
       eprint = {1410.1537},
 primaryClass = {astro-ph.GA},
       adsurl = {https://ui.adsabs.harvard.edu/abs/2015ApJ...802...99K}
}

@article{KiminkiSmith2018,
       author = {{Kiminki}, Megan M. and {Smith}, Nathan},
        title = "{A radial velocity survey of the Carina Nebula's O-type stars}",
      journal = {\mnras},
         year = 2018,
        month = jun,
       volume = {477},
       number = {2},
        pages = {2068-2086},
          doi = {10.1093/mnras/sty748},
archivePrefix = {arXiv},
       eprint = {1803.07057},
 primaryClass = {astro-ph.SR},
       adsurl = {https://ui.adsabs.harvard.edu/abs/2018MNRAS.477.2068K}
}

@article{Kounkel2017,
      author = {{Kounkel}, Marina and {Hartmann}, Lee and {Mateo}, Mario and {Bailey}, III, John I.},
        title = "{Kinematics of the Optically Visible YSOs toward the Orion B Molecular Cloud}",
      journal = {\apj},
         year = 2017,
        month = aug,
       volume = {844},
       number = {2},
          eid = {138},
        pages = {138},
          doi = {10.3847/1538-4357/aa7dea},
archivePrefix = {arXiv},
       eprint = {1707.01115},
 primaryClass = {astro-ph.SR},
       adsurl = {https://ui.adsabs.harvard.edu/abs/2017ApJ...844..138K}
}

@article{Kounkel2018,
       author = {{Kounkel}, Marina and {Covey}, Kevin and {Su{\'a}rez}, Genaro and {Rom{\'a}n-Z{\'u}{\~n}iga}, Carlos and {Hernandez}, Jesus and {Stassun}, Keivan and {Jaehnig}, Karl O. and {Feigelson}, Eric D. and {Pe{\~n}a Ram{\'\i}rez}, Karla and {Roman-Lopes}, Alexandre and {Da Rio}, Nicola and {Stringfellow}, Guy S. and {Kim}, J. Serena and {Borissova}, Jura and {Fern{\'a}ndez-Trincado}, Jos{\'e} G. and {Burgasser}, Adam and {Garc{\'\i}a-Hern{\'a}ndez}, D.~A. and {Zamora}, Olga and {Pan}, Kaike and {Nitschelm}, Christian},
        title = "{The APOGEE-2 Survey of the Orion Star-forming Complex. II. Six-dimensional Structure}",
      journal = {\aj},
         year = 2018,
        month = sep,
       volume = {156},
       number = {3},
          eid = {84},
        pages = {84},
          doi = {10.3847/1538-3881/aad1f1},
archivePrefix = {arXiv},
       eprint = {1805.04649},
 primaryClass = {astro-ph.SR},
       adsurl = {https://ui.adsabs.harvard.edu/abs/2018AJ....156...84K}
}

@article{Kuhn2019,
       author = {{Kuhn}, Michael A. and {Hillenbrand}, Lynne A. and {Sills}, Alison and {Feigelson}, Eric D. and {Getman}, Konstantin V.},
        title = "{Kinematics in Young Star Clusters and Associations with Gaia DR2}",
      journal = {\apj},
         year = 2019,
        month = jan,
       volume = {870},
       number = {1},
          eid = {32},
        pages = {32},
          doi = {10.3847/1538-4357/aaef8c},
archivePrefix = {arXiv},
       eprint = {1807.02115},
 primaryClass = {astro-ph.GA},
       adsurl = {https://ui.adsabs.harvard.edu/abs/2019ApJ...870...32K}
}

@article{Kuhn2020,
      author = {{Kuhn}, Michael A. and {Hillenbrand}, Lynne A. and {Carpenter}, John M. and {Avelar Menendez}, Angel Rodrigo},
        title = "{The Formation of a Stellar Association in the NGC 7000/IC 5070 Complex: Results from Kinematic Analysis of Stars and Gas}",
      journal = {\apj},
         year = 2020,
        month = aug,
       volume = {899},
       number = {2},
          eid = {128},
        pages = {128},
          doi = {10.3847/1538-4357/aba19a},
archivePrefix = {arXiv},
       eprint = {2006.08622},
 primaryClass = {astro-ph.GA},
       adsurl = {https://ui.adsabs.harvard.edu/abs/2020ApJ...899..128K}
}

@article{Kuhn2021,
       author = {{Kuhn}, Michael A. and {de Souza}, Rafael S. and {Krone-Martins}, Alberto and {Castro-Ginard}, Alfred and {Ishida}, Emille E.~O. and {Povich}, Matthew S. and {Hillenbrand}, Lynne A. and {COIN Collaboration}},
        title = "{SPICY: The Spitzer/IRAC Candidate YSO Catalog for the Inner Galactic Midplane}",
      journal = {\apjs},
         year = 2021,
        month = jun,
       volume = {254},
       number = {2},
          eid = {33},
        pages = {33},
          doi = {10.3847/1538-4365/abe465},
archivePrefix = {arXiv},
       eprint = {2011.12961},
 primaryClass = {astro-ph.GA},
       adsurl = {https://ui.adsabs.harvard.edu/abs/2021ApJS..254...33K}
}

@incollection{Lada1987,
       author = {{Lada}, Charles J.},
        title = "{Star formation: from OB associations to protostars.}",
    booktitle = {Star Forming Regions},
         year = 1987,
       editor = {{Peimbert}, Manuel and {Jugaku}, Jun},
       series = {IAU Symposium},
       volume = {115},
        month = jan,
        pages = {1},
        publisher = {D. Reidel Publishing Company},
       adsurl = {https://ui.adsabs.harvard.edu/abs/1987IAUS..115....1L}
}

@article{Ladd2005,
       author = {{Ladd}, Ned and {Purcell}, Cormac and {Wong}, Tony and {Robertson}, Stuart},
        title = "{Beam Size, Shape and Efficiencies for the ATNF Mopra Radio Telescope at 86-115 GHz}",
      journal = {\pasa},
         year = 2005,
        month = jan,
       volume = {22},
       number = {1},
        pages = {62-72},
          doi = {10.1071/AS04068},
       adsurl = {https://ui.adsabs.harvard.edu/abs/2005PASA...22...62L}
}

@article{Lim2023,
       author = {{Lim}, Beomdu and {Hong}, Jongsuk and {Lee}, Jinhee and {Yun}, Hyeong-Sik and {Hwang}, Narae and {Park}, Byeong-Gon},
        title = "{The Kinematics of the Young Stellar Population in the W5 Region of the Cassiopeia OB6 Association: Implication for the Formation Process of Stellar Associations}",
      journal = {\aj},
         year = 2023,
        month = sep,
       volume = {166},
       number = {3},
          eid = {97},
        pages = {97},
          doi = {10.3847/1538-3881/ace772},
archivePrefix = {arXiv},
       eprint = {2307.06557},
 primaryClass = {astro-ph.GA},
       adsurl = {https://ui.adsabs.harvard.edu/abs/2023AJ....166...97L}
}

@article{Lim2025,
       author = {{Lim}, Beomdu and {Yun}, Hyeong-Sik and {Kim}, Hyun-Jeong and {Lee}, Yuna and {Koo}, Jae-Rim and {Hong}, Jongsuk and {Oh}, Heeyoung},
        title = "{Formation Process of the Young Stellar Population in Messier 16 from a Kinematic Perspective}",
      journal = {\aj},
         year = 2025,
        month = oct,
       volume = {170},
       number = {4},
          eid = {228},
        pages = {228},
          doi = {10.3847/1538-3881/adfb60},
archivePrefix = {arXiv},
       eprint = {2508.08666},
 primaryClass = {astro-ph.SR},
       adsurl = {https://ui.adsabs.harvard.edu/abs/2025AJ....170..228L}
}

@article{Lindegren2021,
       author = {{Lindegren}, L. and {Klioner}, S.~A. and {Hern{\'a}ndez}, J. and {Bombrun}, A. and {Ramos-Lerate}, M. and {Steidelm{\"u}ller}, H. and {Bastian}, U. and {Biermann}, M. and {de Torres}, A. and {Gerlach}, E. and {Geyer}, R. and {Hilger}, T. and {Hobbs}, D. and {Lammers}, U. and {McMillan}, P.~J. and {Stephenson}, C.~A. and {Casta{\~n}eda}, J. and {Davidson}, M. and {Fabricius}, C. and {Gracia-Abril}, G. and {Portell}, J. and {Rowell}, N. and {Teyssier}, D. and {Torra}, F. and {Bartolom{\'e}}, S. and {Clotet}, M. and {Garralda}, N. and {Gonz{\'a}lez-Vidal}, J.~J. and {Torra}, J. and {Abbas}, U. and {Altmann}, M. and {Anglada Varela}, E. and {Balaguer-N{\'u}{\~n}ez}, L. and {Balog}, Z. and {Barache}, C. and {Becciani}, U. and {Bernet}, M. and {Bertone}, S. and {Bianchi}, L. and {Bouquillon}, S. and {Brown}, A.~G.~A. and {Bucciarelli}, B. and {Busonero}, D. and {Butkevich}, A.~G. and {Buzzi}, R. and {Cancelliere}, R. and {Carlucci}, T. and {Charlot}, P. and {Cioni}, M.-R.~L. and {Crosta}, M. and {Crowley}, C. and {del Peloso}, E.~F. and {del Pozo}, E. and {Drimmel}, R. and {Esquej}, P. and {Fienga}, A. and {Fraile}, E. and {Gai}, M. and {Garcia-Reinaldos}, M. and {Guerra}, R. and {Hambly}, N.~C. and {Hauser}, M. and {Jan{\ss}en}, K. and {Jordan}, S. and {Kostrzewa-Rutkowska}, Z. and {Lattanzi}, M.~G. and {Liao}, S. and {Licata}, E. and {Lister}, T.~A. and {L{\"o}ffler}, W. and {Marchant}, J.~M. and {Masip}, A. and {Mignard}, F. and {Mints}, A. and {Molina}, D. and {Mora}, A. and {Morbidelli}, R. and {Murphy}, C.~P. and {Pagani}, C. and {Panuzzo}, P. and {Pe{\~n}alosa Esteller}, X. and {Poggio}, E. and {Re Fiorentin}, P. and {Riva}, A. and {Sagrist{\`a} Sell{\'e}s}, A. and {Sanchez Gimenez}, V. and {Sarasso}, M. and {Sciacca}, E. and {Siddiqui}, H.~I. and {Smart}, R.~L. and {Souami}, D. and {Spagna}, A. and {Steele}, I.~A. and {Taris}, F. and {Utrilla}, E. and {van Reeven}, W. and {Vecchiato}, A.},
        title = "{Gaia Early Data Release 3. The astrometric solution}",
      journal = {\aap},
         year = 2021,
        month = may,
       volume = {649},
          eid = {A2},
        pages = {A2},
          doi = {10.1051/0004-6361/202039709},
archivePrefix = {arXiv},
       eprint = {2012.03380},
 primaryClass = {astro-ph.IM},
       adsurl = {https://ui.adsabs.harvard.edu/abs/2021A&A...649A...2L}
}

@article{Liu2024,
       author = {{Liu}, Dejian and {Xu}, Ye and {Li}, YingJie and {Lin}, Zehao and {Hao}, Chaojie and {Yang}, WenJin and {Li}, Jingjing and {Liu}, Xinrong and {Dong}, Yiwei and {Bian}, Shuaibo and {Kong}, Deyun},
        title = "{Molecular Bubble and Outflow in S Mon Revealed by Multiband Data Sets}",
      journal = {\apj},
         year = 2024,
        month = mar,
       volume = {964},
       number = {1},
          eid = {93},
        pages = {93},
          doi = {10.3847/1538-4357/ad24e0},
archivePrefix = {arXiv},
       eprint = {2401.17525},
 primaryClass = {astro-ph.GA},
       adsurl = {https://ui.adsabs.harvard.edu/abs/2024ApJ...964...93L}
}

@article{LoktinPopov2020,
       author = {{Loktin}, Alexander V. and {Popov}, Alexander A.},
        title = "{Rotation of open clusters based on Gaia data. Praesepe}",
      journal = {Astronomische Nachrichten},
         year = 2020,
        month = jul,
       volume = {341},
       number = {638},
        pages = {638-644},
          doi = {10.1002/asna.202013687},
       adsurl = {https://ui.adsabs.harvard.edu/abs/2020AN....341..638L}
}

@article{Majewski2017,
       author = {{Majewski}, Steven R. and {Schiavon}, Ricardo P. and {Frinchaboy}, Peter M. and {Allende Prieto}, Carlos and {Barkhouser}, Robert and {Bizyaev}, Dmitry and {Blank}, Basil and {Brunner}, Sophia and {Burton}, Adam and {Carrera}, Ricardo and {Chojnowski}, S. Drew and {Cunha}, K{\'a}tia and {Epstein}, Courtney and {Fitzgerald}, Greg and {Garc{\'\i}a P{\'e}rez}, Ana E. and {Hearty}, Fred R. and {Henderson}, Chuck and {Holtzman}, Jon A. and {Johnson}, Jennifer A. and {Lam}, Charles R. and {Lawler}, James E. and {Maseman}, Paul and {M{\'e}sz{\'a}ros}, Szabolcs and {Nelson}, Matthew and {Nguyen}, Duy Coung and {Nidever}, David L. and {Pinsonneault}, Marc and {Shetrone}, Matthew and {Smee}, Stephen and {Smith}, Verne V. and {Stolberg}, Todd and {Skrutskie}, Michael F. and {Walker}, Eric and {Wilson}, John C. and {Zasowski}, Gail and {Anders}, Friedrich and {Basu}, Sarbani and {Beland}, Stephane and {Blanton}, Michael R. and {Bovy}, Jo and {Brownstein}, Joel R. and {Carlberg}, Joleen and {Chaplin}, William and {Chiappini}, Cristina and {Eisenstein}, Daniel J. and {Elsworth}, Yvonne and {Feuillet}, Diane and {Fleming}, Scott W. and {Galbraith-Frew}, Jessica and {Garc{\'\i}a}, Rafael A. and {Garc{\'\i}a-Hern{\'a}ndez}, D. An{\'\i}bal and {Gillespie}, Bruce A. and {Girardi}, L{\'e}o and {Gunn}, James E. and {Hasselquist}, Sten and {Hayden}, Michael R. and {Hekker}, Saskia and {Ivans}, Inese and {Kinemuchi}, Karen and {Klaene}, Mark and {Mahadevan}, Suvrath and {Mathur}, Savita and {Mosser}, Beno{\^\i}t and {Muna}, Demitri and {Munn}, Jeffrey A. and {Nichol}, Robert C. and {O'Connell}, Robert W. and {Parejko}, John K. and {Robin}, A.~C. and {Rocha-Pinto}, Helio and {Schultheis}, Matthias and {Serenelli}, Aldo M. and {Shane}, Neville and {Silva Aguirre}, Victor and {Sobeck}, Jennifer S. and {Thompson}, Benjamin and {Troup}, Nicholas W. and {Weinberg}, David H. and {Zamora}, Olga},
        title = "{The Apache Point Observatory Galactic Evolution Experiment (APOGEE)}",
      journal = {\aj},
         year = 2017,
        month = sep,
       volume = {154},
       number = {3},
          eid = {94},
        pages = {94},
          doi = {10.3847/1538-3881/aa784d},
archivePrefix = {arXiv},
       eprint = {1509.05420},
 primaryClass = {astro-ph.IM},
       adsurl = {https://ui.adsabs.harvard.edu/abs/2017AJ....154...94M}
}

@article{Megeath2022,
       author = {{Megeath}, S.~T. and {Gutermuth}, R.~A. and {Kounkel}, M.~A.},
        title = "{Low Mass Stars as Tracers of Star and Cluster Formation}",
      journal = {\pasp},
         year = 2022,
        month = apr,
       volume = {134},
       number = {1034},
          eid = {042001},
        pages = {042001},
          doi = {10.1088/1538-3873/ac4c9c},
archivePrefix = {arXiv},
       eprint = {2203.03655},
 primaryClass = {astro-ph.GA},
       adsurl = {https://ui.adsabs.harvard.edu/abs/2022PASP..134d2001M}
}

@article{MelnikDambis2020,
       author = {{Melnik}, A.~M. and {Dambis}, A.~K.},
        title = "{Internal motions in OB associations with Gaia DR2}",
      journal = {\mnras},
         year = 2020,
        month = apr,
       volume = {493},
       number = {2},
        pages = {2339-2351},
          doi = {10.1093/mnras/staa454},
archivePrefix = {arXiv},
       eprint = {2002.05044},
 primaryClass = {astro-ph.GA},
       adsurl = {https://ui.adsabs.harvard.edu/abs/2020MNRAS.493.2339M}
}

@article{Nurnberger2002,
       author = {{N{\"u}rnberger}, D.~E.~A. and {Bronfman}, L. and {Yorke}, H.~W. and {Zinnecker}, H.},
        title = "{The molecular environment of NGC 3603. I. Spatial distribution and kinematic structure}",
      journal = {\aap},
         year = 2002,
        month = oct,
       volume = {394},
        pages = {253-269},
          doi = {10.1051/0004-6361:20021022},
       adsurl = {https://ui.adsabs.harvard.edu/abs/2002A&A...394..253N}
}

@article{Offner2008,
       author = {{Offner}, Stella S.~R. and {Krumholz}, Mark R. and {Klein}, Richard I. and {McKee}, Christopher F.},
        title = "{The Kinematics of Molecular Cloud Cores in the Presence of Driven and Decaying Turbulence: Comparisons with Observations}",
      journal = {\aj},
         year = 2008,
        month = jul,
       volume = {136},
       number = {1},
        pages = {404-420},
          doi = {10.1088/0004-6256/136/1/404},
archivePrefix = {arXiv},
       eprint = {0712.3053},
 primaryClass = {astro-ph},
       adsurl = {https://ui.adsabs.harvard.edu/abs/2008AJ....136..404O}
}

@article{Povich2011,
       author = {{Povich}, Matthew S. and {Smith}, Nathan and {Majewski}, Steven R. and {Getman}, Konstantin V. and {Townsley}, Leisa K. and {Babler}, Brian L. and {Broos}, Patrick S. and {Indebetouw}, R{\'e}my and {Meade}, Marilyn R. and {Robitaille}, Thomas P. and {Stassun}, Keivan G. and {Whitney}, Barbara A. and {Yonekura}, Yoshinori and {Fukui}, Yasuo},
        title = "{A Pan-Carina Young Stellar Object Catalog: Intermediate-mass Young Stellar Objects in the Carina Nebula Identified Via Mid-infrared Excess Emission}",
      journal = {\apjs},
         year = 2011,
        month = may,
       volume = {194},
       number = {1},
          eid = {14},
        pages = {14},
          doi = {10.1088/0067-0049/194/1/14},
archivePrefix = {arXiv},
       eprint = {1103.2060},
 primaryClass = {astro-ph.SR},
       adsurl = {https://ui.adsabs.harvard.edu/abs/2011ApJS..194...14P}
}

@article{Preibisch2012,
       author = {{Preibisch}, T. and {Roccatagliata}, V. and {Gaczkowski}, B. and {Ratzka}, T.},
        title = "{Herschel far-infrared observations of the Carina Nebula complex. I. Introduction and global cloud structure}",
      journal = {\aap},
         year = 2012,
        month = may,
       volume = {541},
          eid = {A132},
        pages = {A132},
          doi = {10.1051/0004-6361/201218851},
archivePrefix = {arXiv},
       eprint = {1204.0627},
 primaryClass = {astro-ph.GA},
       adsurl = {https://ui.adsabs.harvard.edu/abs/2012A&A...541A.132P}
}
\bibliographystyle{aasjournalv7}



\end{document}